\documentclass{jfm}
\usepackage{makeidx}
\usepackage{amssymb}
\usepackage{graphicx}
\usepackage{newtxtext}
\usepackage{newtxmath}
\usepackage{natbib}
\usepackage{hyperref}
\usepackage{subfigure}

\hypersetup{
    colorlinks = true,
    urlcolor   = blue,
    citecolor  = black,
}

\newcommand{\RomanNumeralCaps}[1]
\linenumbers
\affiliation{\aff{1}Department of Engineering Mechanics, Tsinghua University, Beijing 100084, China}
\begin{document}

\title{The dual solution stability gap bounded by sub- and supercritical
geometric thresholds in steady shock reflection}
\author{Xue-Ying Wang and Zi-Niu Wu \aff{1} \corresp{%
\email{ziniuwu@tsinghua.edu.cn}},}
\maketitle

\begin{abstract}
In this paper, we study the lower limit of geometric setup (expressed in
terms of the relative trailing edge height) for stable shock reflection
within the dual solution domain, where both regular reflection (RR) and Mach
reflection (MR) are theoretically possible. We prove that the lower limit
for MR is larger than that for RR in the dual solution domain, and this
proof relies on the use of minimum Mach stem height that can be evaluated
exactly. We thus identify two critical thresholds (expressed in terms of the
relative trailing edge height): a subcritical threshold, below which both
reflection modes are possibly unstable, and a supercritical threshold, above
which both become stable. The mismatch between these two thresholds gives
rise to a dual solution stability gap---a range of geometric configurations
where RR remains stable while MR is unstable. This implies that, within this
gap, a steady RR solution (start flow) may undergo a dynamic transition to a
possibly unsteady or unstable MR configuration (unstart flow) under
sufficiently large upstream or localized disturbances. We verify the
existence of this stability gap, both theoretically and numerically, and
demonstrate the time history of the associated dynamic transition through
numerical simulations. Complex flow structures, such as hybrid MR --- type
VI shock interference, and double MR --- MR, are found to exist during the
dynamic transition. Apart from direct dynamic transition from RR to MR to
unstart flow, we also observe inverted dynamic transition, for which RR
transits to MR but then transit back to RR.
\end{abstract}

\section{Introduction}

\begin{figure} 
\begin{center}
\subfigure a) \  \includegraphics[width=0.38\textwidth]{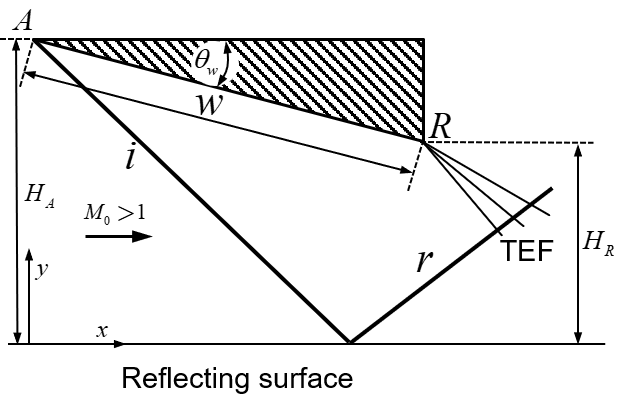} \subfigure %
b) \  \includegraphics[width=0.38\textwidth]{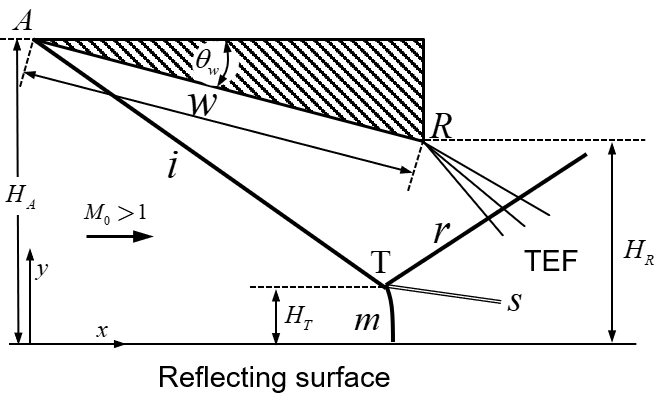}
\end{center}
\caption{Illustration of shock reflection. a) RR, b) MR.}
\label{fig-xw-1}
\end{figure}

The reflection of an incident shock over a reflecting surface (symmetric
line in case of symmetric reflection) is an important phenomenon in steady
supersonic flow and has received considerable amount of studies since 1970s
(Ben-Dor 2007). The critical condition at which regular reflection (RR) or
Mach reflection (MR) (as illustrated in Figure \ref{fig-xw-1}) occurs is one
of the important issues that have been studied.

There are two classical transition criteria, the detachment condition and
the von Neumann condition (von Neumann 1943). Consider the configuration
shown in Figure \ref{fig-xw-1}. For a given inflow Mach number (denoted $%
M_{0}$ in this paper), if the wedge angle (denoted $\theta _{w}$ in this
paper), i.e., the deflection angle of the wedge generating the incident
shock wave (i), is larger than the detachment condition (denoted $\theta
_{w}^{(D)}(M_{0})$), then we necessarily have Mach reflection. If the wedge
angle is smaller than the von Neumann condition (denoted $\theta
_{w}^{(N)}(M_{0})$), then we necessarily have regular reflection. Note that,
inverted Mach reflection that corresponds to Mach reflection below the von
Neumann condition (Henderson \& Lozzi 1979; Hornung 1986; Hekiri \& Emanuel
2015) could happen when there are additional influences such as downstream
body (Roye, Henderson \& Menikoff 1998), high downstream pressure (Ben-Dor
et al. 1999), asymmetry shock reflection (Li, Chpoun, Ben-Dor, 1999), a
downstream incident shock (Guan, Bai \& Wu 2018), but this inverted Mach
reflection will not be considered here.

There is a dual solution domain in the plane $(M_{0},\theta _{w})$, since
the detachment condition ($\theta _{w}$ $=\theta _{w}^{(D)}(M_{0})$) is
above the von Neumann condition ($\theta _{w}$ $=\theta _{w}^{(N)}(M_{0})$).
This means that when $\theta _{w}^{(N)}(M_{0})<\theta _{w}$ $<\theta
_{w}^{(D)}(M_{0})$, both regular reflection and Mach reflection are
possible. Though such dual solution was recognized to be theoretically
possible, great effort has been required to clarify the real transition
process.

Early studies only observed Mach reflection (Henderson\&Lozzi 1975; Hornung,
Oertel \& Sandeman 1979). The failure to observe regular reflection led
Hornung \&Robinson (1982) to conjecture that regular reflection is unstable
in the dual solution domain. However, the stability analysis conducted by
Teshukov (1989) and Li \& Ben-Dor (1996) proved that regular reflection
should be stable. In fact, Hornung, Oertel \& Sandeman (1979) hypothesized a
hysteresis process, which means that, in the dual solution domain, whether
we have Mach reflection or regular reflection depends on the history of the
building of the actual steady flow. Then regular reflection was observed
numerically (Vuillon, Zeitoun\&Ben-Dor 1995) and experimentally (Chpoun et
al. 1995), where hysteresis occurs by changing the wedge angle. Later on
similar hysteresis was observed when the inflow Mach number changes from
different directions (Ivanov et al 2001). More discussions about the
hysteresis can be found in the paper of Ben-Dor et al. (2002) and Hornung
(2014).

Apart from transition at $\theta _{w}$ crossing the boundaries of the dual
solution domain, dynamic transition within the dual solution domain is also
possible. For instance, RR may transit to MR if the amplitude of the
disturbance exceeds a certain level, according to a number of studies
(Ivanov et al. 1997, 1998, 2000; Kudryavtsev et al. 2002; Li, Gao \& Wu
2011).

In the 1990s, it was discovered that stable Mach reflection is also
subjected to geometric constraint. Vuillon, Zeitoun \& Ben-Dor (1995)
studied shock reflection for extremely lower and larger values of the
trailing edge height $H_{R}$ (the distance of the trailing edge (R) of the
wedge from the reflecting surface, see Figure \ref{fig-xw-1}). Based on
theoretical considerations, they found that the distance $H_{R}$ is bounded
by a lower limit $H_{R,\min }$ and an upper limit $H_{R,\max }$. The lower
limit corresponds to the case in which the reflected shock wave (r) grazes
the trailing edge (R). \ The upper limit is determined by the point where
the leading characteristic of the trailing edge expansion fan (denoted TEF
in Figure \ref{fig-xw-1}) intersects the incident shock wave at the
reflection point (G) for regular reflection and at the triple point (T) for
Mach reflection. More works have been given by Li \& Ben-Dor (1997) and
Grasso \& Paoli (1999), the latter studied the effect of geometric set-up
when accounting for shock reflection in nonequilibrium flow.

The significance of upper limit has been more studied than the lower limit,
since beyond the upper limit occurs another interesting transition
phenomenon: transition from MR to RR. Vuillon, Zeitoun \& Ben-Dor (1995)
argued that increasing $H_{R}$ may trigger transition from MR to RR and
believed this transition may occur for $H_{R}<H_{R,\max }$. Later on Li \&
Ben-Dor (1997) revisted the expression of $H_{R,\max }$ and clarified that
MR to RR transition occurs for $H_{R}>H_{R,\max }$, i.e., beyond the
threshold at which interaction between the trailing edge expansion fan and
incident shock occurs. Bai (2023) then provided a new critical condition
(corresponding to a $H_{R}$ larger than $H_{R,\max }$) at which transition
from Mach reflection to regular reflection occurs, and found that beyond the
upper limit the Mach stem height decreases nonlinearly with the trailing
edge height, until it vanishes at this new critical condition.

The lower limit, though less studies, is also of great significance. \
According to Vuillon, Zeitoun \& Ben-Dor (1995), whenever the distance $%
H_{R} $ reaches or is reduced below $H_{R,\min }$, the Mach reflection
becomes unstable and its Mach stem moves upstream until the Mach reflection
vanishes and a bow shock \ wave is established ahead of the leading edge of
the reflecting wedge. As a result, the flow through the two-dimensional
converging nozzle, formed by the surface of the reflecting wedge and the
line of symmetry, becomes subsonic. The two-dimensional converging nozzle
which is formed by the wedge and bottom surfaces is said to be unstarted or
choked. This process was indeed observed not only by their numerical
simulations but also in experiments (Chpoun et al. 1995).

There is a great difficult to study the influence of $H_{R,\min }$ for Mach
reflection, since the expression for $H_{R,\min }$ involves the unknown Mach
stem height $H_{T}$ (see Figure \ref{fig-xw-1}(b)). Vuillon, Zeitoun \&
Ben-Dor (1995) suggested to use the model of Azevedo \& Liu (1993). Li \&
Ben-Dor (1997) derived their own model for $H_{T}$ when studying the
geometric set-up. These earlier models and more recent models such as the
models by Mouton \& Hornung (2008), Gao \& Wu (2010) and Bai \& Wu (2017)
are all approximative, so it is impossible to find the exact value of $%
H_{R,\min }$ for Mach reflection.

Using their own Mach stem height model, Li \& BenDor (1997) found that, with
the particular condition $M_{0}=5$ within the dual solution domain, $%
H_{R,\min }^{(MR)}$ ($H_{R,\min }$ for Mach reflection) is larger than $%
H_{R,\min }^{(RR)}$ ($H_{R,\min }$ for regular reflection). It is unknown
that this conclusion holds exactly in the dual solution domain. If this is
indeed so, then it possibly implies a new transition scenario: for $%
H_{R,\min }^{(RR)}<H_{R}$ $<H_{R,\min }^{(MR)}$, regular reflection is
stable and Mach reflection is unstable in the sense that the flow will be
unstarted or chocked as pointed out by Vuillon, Zeitoun \& Ben-Dor (1995).
More interestingly, if such a stable regular reflection is subjected to
large amplitude disturbance, then this stable regular reflection may transit
dynamically to unstable MR, a phenomenon that appears to have not been
observed before. The confirmation of the difference between $H_{R,\min
}^{(RR)}$ and $H_{R,\min }^{(MR)}$ in the dual solution domain, and the
discussion of the significance of this difference in shock transition is the
object of the present study.

In section 2, we express the expressions of $H_{R,\min }^{(RR)}$ and $%
H_{R,\min }^{(MR)}$ in terms of $M_{0}$ and $\theta _{w}$, and\ prove that
the inequality $H_{R,\min }^{(RR)}<H_{R,\min }^{(MR)}$ holds for $M_{0}$ and
$\theta _{w}$ within the dual solution domain. The significance of this
inequality in identifying a dual solution stablity gap will be discussed.

In section 3, we look at how large will be the magnitude of the dual
solution stability gap. Since this gap depends on the Mach stem height, and
the existing Mach stem height model is not accurate enough to provide
quantitatively exact values of the stability gap, we will use a linear Mach
stem height model where the coefficients come from high fidelity numerical
simulation, to give very exact estimation of the stability gap for some
particular condition.

In section 4, we use numerical simulation to study dynamic transition from
stable RR and unstable MR, for $H_{R}$ lying inside the dual solution
stability gap $H_{R,\min }^{(RR)}<H_{R}<H_{R,\min }^{(MR)}$. This not only
provides further evidence for the existence of dual solution stability gap,
but also shows how the transition evolves dynamically and depends on the
upstream disturbance. Such a study also complements the previous studies
about dynamic transition from stable RR to stable MR, for the condition $%
H_{R}>H_{R,\min }^{(MR)}$ (though this condition was implicitly assumed).

\section{The subcritical and supercritical geometric thresholds and dual
solution stability gap}

The purpose of this section is to \ prove that the inequality $H_{R,\min
}^{(RR)}<H_{R,\min }^{(MR)}$ holds for $M_{0}$ and $\theta _{w}$ within the
dual solution domain, and to introduce the notions of subcritical geometric
threshhold, supercritical geometric threshhold, and dual solutuon stability
gap.

\subsection{Basic expression for the lower limit of the geometric setup}

\begin{figure} 
\begin{center}
\subfigure a) \  \includegraphics[width=0.38\textwidth]{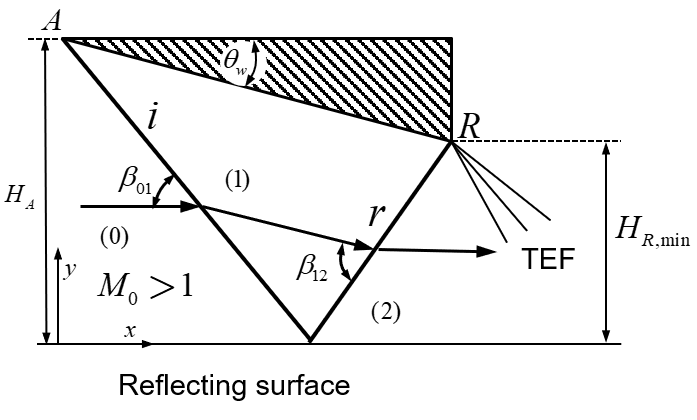} \subfigure %
b) \  \includegraphics[width=0.38\textwidth]{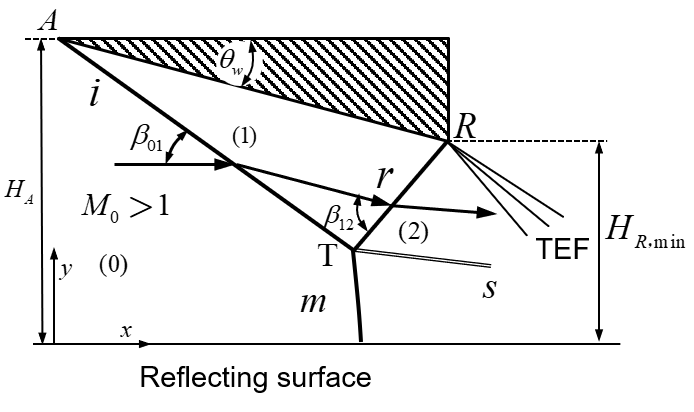}
\end{center}
\caption{Shock reflection for $H_{R}=H_{R,\min }$ at which the reflecting
shock intersects the trailing edge $R$. a) RR, b) MR.}
\label{fig-1}
\end{figure}

We recall that the lower limit $H_{R,\min }$ is the value of $H_{R}$ at
which the reflected shock grazes the trailing edge (Vuillon, Zeitoun \&
Ben-Dor 1995). The geometrical set-up for $H_{R}=H_{R,\min }$ is
schematically displayed in Figure \ref{fig-1}(a) for RR and Figure \ref%
{fig-1}(b) for MR.

From the geometric relations displayed in Figure \ref{fig-1}(b), it can be
shown that the lower limit $H_{R,\min }$ is related to the Mach stem height $%
H_{T}$ by

\begin{equation}
H_{R,\min }=H_{T}+\phi w  \label{xy-2-1-1}
\end{equation}%
where\ $w$ is the length of the wedge lower surface, and $\phi $ is given by
\begin{equation}
\phi =\frac{\tan \beta _{01}\tan (\beta _{12}-\theta _{w})\cos \theta
_{w}-\tan (\beta _{12}-\theta _{w})\sin \theta _{w}}{\tan \beta _{01}+\tan
(\beta _{12}-\theta _{w})},  \label{xy-2-1-3}
\end{equation}%
or equivalently by

\[
\phi =\left \{
\begin{array}{l}
\frac{\sin \beta _{01}\sin \beta _{12}}{\sin (\beta _{01}+\beta _{12}-\theta
_{w})}-\sin \theta _{w}\text{ (Vuillon, Zeitoun \& Ben-Dor 1995)} \\
\frac{\sin (\beta _{12}-\theta _{w})\sin (\beta _{01}-\theta _{w})}{\sin
(\beta _{01}+\beta _{12}-\theta _{w})}\text{ (Li \& Ben-Dor 1997)}%
\end{array}%
\right. .
\]

In (\ref{xy-2-1-3}), $\beta _{01}$ is the shock angle of the incident shock
(i), $\beta _{12}$ is the shock angle of the reflected shock (in the
vicinity of the triple point for MR or of the reflection point for RR).
These parameters are determined by the von Neumann two shock (for RR) or
three shock theories (for MR) (Vuillon, Zeitoun \& Ben-Dor 1995).

For regular reflection, we have $H_{T}=0$ in equation (\ref{xy-2-1-1}). For
Mach reflection, the expression (\ref{xy-2-1-1}) for $H_{R,\min }$ depends
on the Mach stem height $H_{T}$ that is also an unknown. In the next we will
introduce the functional form of $H_{T}$ to show so that the relative value
of this lower limit only depends on\ $M_{0}$ and $\theta _{w}$.

\subsection{Equivalent form of the lower limit that depends on $M_{0}$ and $%
\protect \theta _{w}$ only}

The expression (\ref{xy-2-1-1}) and (\ref{xy-2-1-3}) depend on several
parameters including $H_{T}$ and $w$. In order to study the problem in the
dual solution domain (in the plane $M_{0}$ and $\theta _{w}$), it is better
to make (\ref{xy-2-1-1}) and (\ref{xy-2-1-3}) to be equivalent to a form
that depends on $M_{0}$ and $\theta _{w}$ only.

The wedge lower surface length $w$ can be related to $\theta _{w}$ by the
obvious geometric relation
\begin{equation}
w=\frac{H_{A}-H_{R}}{\sin \theta _{w}}.  \label{xy-2-2-1}
\end{equation}%
The Mach stem height $H_{T}$ is apparently an unknown parameter. In order to
prove that the lower limit depends on $M_{0}$ and $\theta _{w}$ only, we
just need to use the functional form of $H_{T}$. Hornung \& Robinson (1982)
argued that the Mach stem height is affected by the pressure decreasing
information from the wedge trailing edge expansion fan so the Mach stem
height must follow the functional form given by
\[
\frac{H_{T}}{w}=h\left( M_{0},\theta _{w},\frac{H_{R}}{w}\right)
\]%
which can be rewritten as
\begin{equation}
\frac{H_{T}}{H_{A}}=f\left( M_{0},\theta _{w},\frac{H_{R}}{H_{A}}\right)
\label{xy-2-2-1c}
\end{equation}%
if (\ref{xy-2-2-1}) for $w$ is used. Note that the functional form also
depends on $\gamma $, the ratio of the specific ratios. Here we do not
consider varying $\gamma $ so this parameter is considered as a constant in (%
\ref{xy-2-2-1c}).

In the following we will express the results in terms of the relative
trailing edge height $g$ defined by%
\[
g=\frac{H_{R}}{H_{A}}.
\]%
Accordingly, the lower limit of the relative trailing edge height is denoted
as
\begin{equation}
g_{\min }=\frac{H_{R,\min }}{H_{A}}.  \label{eq-4-last}
\end{equation}

Putting $H_{R}=H_{R,\min }$ in (\ref{xy-2-2-1}) and (\ref{xy-2-2-1c}), and
using (\ref{eq-4-last}), it can be proven that the lower limit expression (%
\ref{xy-2-1-1}) becomes%
\begin{equation}
g_{\min }=f\left( M_{0},\theta _{w},g_{\min }\right) +\frac{\phi }{\sin
\theta _{w}}\left( 1-g_{\min }\right) .  \label{eq-xy-main-1}
\end{equation}%
The expression (\ref{eq-xy-main-1}) means that $g_{\min }$ \ indeed depends
on $M_{0}$ and $\theta _{w}$ only.

As stated in Introduction, $g_{\min }^{(RR)}$ denotes $g_{\min }$ for RR and
$g_{\min }^{(MR)}$ denotes $g_{\min }$ for MR. For regular reflection, the
expression (\ref{eq-xy-main-1}) is applied with $f=0$, so
\begin{equation}
g_{\min }^{(RR)}=\frac{\phi ^{RR}}{\sin \theta _{w}+\phi ^{RR}}
\label{eq-4-2-11}
\end{equation}%
with%
\[
\phi ^{RR}=\frac{{\tan {\beta _{01}}\tan \beta _{_{12}}^{RR}\cos {\theta _{w}%
}+\tan {\beta _{01}}\sin {\theta _{w}}}}{{\left( {\tan {\beta _{01}}+\tan
\beta _{_{12}}^{RR}}\right) \sin {\theta _{w}}}}.
\]

\subsection{Proof of the phenomenon of lower limit difference}

The lower limit difference ($g_{\min }^{(MR)}>g_{\min }^{(RR)}$) in the dual
solution domain, as mentioned in Introduction, is to be proven here. As will
be stated in Section 2.4, the existence of this difference will lead to the
notions of subcritical and supercritical geometric thresholds, and \ will
justify the existence of dual solution stability gap.

Now let us prove $g_{\min }^{(MR)}>g_{\min }^{(RR)}$, or equivalently $%
H_{R,\min }^{(MR)}>H_{R,\min }^{(RR)}$, in the dual solution domain.

\begin{figure} 
\begin{center}
\subfigure a) \  \includegraphics[width=0.38\textwidth]{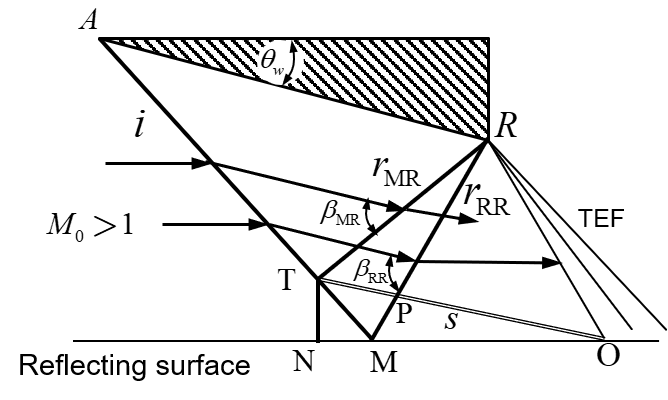} \subfigure %
b) \  \includegraphics[width=0.38\textwidth]{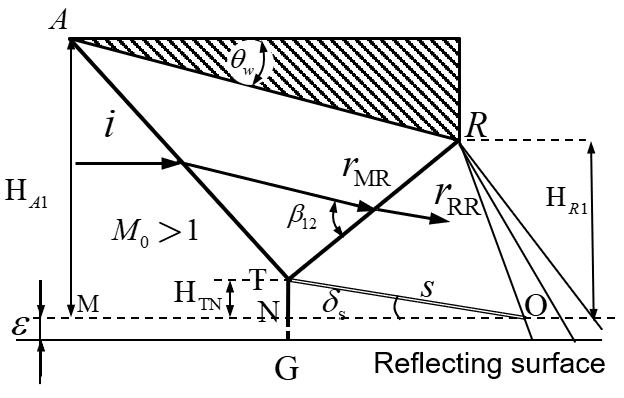}
\end{center}
\caption{Schematic display of lower limit setup. a) RR and MR supposed to
have the same $H_{R}=H_{R,\min }$. b) MR at $H_{R}=H_{R,\min }$.}
\label{fig4-1}
\end{figure}

One may argue that $H_{R,\min }^{(MR)}$ is obviously greater than $H_{R,\min
}^{(RR)}$ in the dual solution domain because the existence of Mach stem
elevates the reflected shock, so that it can graze the trailing edge
earlier. Li \& Ben-Dor (1997) indeed stated that it is evident that the
minimum value of $H_{R}$ for RR is smaller than that for MR. However, as
shown in Figure \ref{fig4-1} (a), where we have superimposed reflected shock
waves for both RR and MR, the reflected shock for Mach reflection and the
reflected shock for regular reflection have different shock angles. The
shock angle $\beta _{12}^{MR}$ for Mach reflection is smaller than the shock
angle\ $\beta _{12}^{RR}$ for regular reflection. It is unclear whether the
elevation of the reflected shock through the Mach stem can compensate the
counter-acting effect of the reduction of the shock angle.

Apparently, there is a great difficult to prove $g_{\min }^{(MR)}>g_{\min
}^{(RR)}$, since $g_{\min }^{(MR)}$ depends on the relative Mach stem height
$f=H_{T}/H_{A}$ which is an unknown. However, as stated in Introduction,
exactly evaluating $H_{T}$ is still challenging.

To overcome this difficulty we will choose to prove $g_{\min
}^{(MR)}>g_{\min }^{(N)}>g_{\min }^{(RR)}$ where $g_{\min }^{(N)}$ is $%
g_{\min }^{(MR)}$\ for minimum Mach stem height $H_{TN}$ (which can be given
exactly). The minimum Mach stem height $H_{TN}$ is defined in Figure \ref%
{fig4-1}(b), which shows Mach reflection at $H_{R}=H_{R,\min }$. It is the
distance between the triple point (T) and the horizontal line (NO), where O
is the intersection point between the initial segment of the slipline (s)
and the leading characteristic line of the wedge trailing edge expansion
fan. The total Mach stem height satisfies%
\begin{equation}
H_{T}=H_{TN}+\varepsilon  \label{eq-HT-epsilon}
\end{equation}%
where $\varepsilon $, with $\varepsilon >0$, is the additional height due to
the passage of the flow across the flow duct below intersection point O.

The quantity $g_{\min }^{(N)}$ is defined by
\[
g_{\min }^{(N)}=\frac{{{H_{R1}}}}{{{H_{A1}}}}
\]%
where $H_{A1}$ is the inlet height above the line NO, $H_{R1}$ is the
trailing edge height above NO, see Figure \ref{fig4-1}(b). Now we have
\begin{equation}
g_{\min }^{MR}=\frac{{{H_{R1}}+\varepsilon }}{{{H_{A1}}+\varepsilon }}
\label{eq-6-1}
\end{equation}

Since $\frac{{{H_{R1}}+\varepsilon }}{{{H_{A1}}+\varepsilon }}>\frac{{{H_{R1}%
}}}{{{H_{A1}}}}$ for any $\varepsilon >0$,\ the following inequality holds
\[
g_{\min }^{MR}>g_{\min }^{(N)}
\]%
Thus, if we can prove $g_{\min }^{(N)}>g_{\min }^{(RR)}$ , then we
necessarily have $g_{\min }^{(MR)}>g_{\min }^{(RR)}$.

Now we prove $g_{\min }^{(N)}>g_{\min }^{(RR)}$. Using the geometric
relations displayed in Figure \ref{fig4-1}(b), it can be shown that, the
condition that the reflected shock grazes the trailing edge leads to
\begin{equation}
\frac{{{H_{A1}}-}H_{TN}}{{\tan {\beta _{01}}}}+\frac{H_{R1}{-{H_{TN}}}}{{%
\tan (\beta _{12}^{MR}-{\theta _{w}})}}=\frac{{{H_{A1}}-{H_{R1}}}}{{\tan {%
\theta _{w}}}}  \label{eq-5-1a}
\end{equation}%
\ The total length of the horizontal line MO, where M is the intersection of
the line NO and the vertical line passing the vertex A, is equal to the
wedge horizontal length plus the horizontal projection of the leading
characteristic line RO, so

\begin{equation}
\frac{{{H_{A1}}-{H_{T1}}}}{{\tan {\beta _{01}}}}+\frac{{{H_{T1}}}}{{\tan {%
\delta _{s}}}}=\frac{{{H_{A1}}-{H_{R1}}}}{{\tan {\theta _{w}}}}+\frac{{{%
H_{R1}}}}{{\tan ({\mu _{1}}+{\delta _{s}})}}  \label{eq-5-1b}
\end{equation}

Solving (\ref{eq-5-1a}) and (\ref{eq-5-1b}) for $g_{\min }^{(N)}=\frac{{{%
H_{R1}}}}{{{H_{A1}}}}$ gives%
\begin{equation}
g_{\min }^{(N)}=\frac{{\frac{\phi ^{MR}{{t_{2}}}}{{\sin {\theta _{w}}}}+%
\frac{{{t_{1}}}}{{\tan {\theta _{w}}}}-\tan {\delta _{s}}}}{{{t_{2}}+\frac{%
\phi ^{MR}{{t_{2}}}}{{\sin {\theta _{w}}}}+\frac{{{t_{1}}}}{{\tan {\theta
_{w}}}}-\frac{{{t_{1}}}}{{\tan ({\mu _{1}}+{\delta _{s}})}}}}  \label{eq-hr1}
\end{equation}%
where ${t_{1}}=\tan {\beta _{01}}\tan {\delta _{s}}$, ${t_{2}}=\tan {\beta
_{01}}-\tan {\delta _{s}}$ and
\[
\phi ^{MR}=\frac{{\tan {\beta _{01}}\tan \beta _{_{12}}^{MR}\cos {\theta _{w}%
}+\tan {\beta _{01}}\sin {\theta _{w}}}}{{\left( {\tan {\beta _{01}}+\tan
\beta _{_{12}}^{MR}}\right) \sin {\theta _{w}}}}
\]

Now we display the quantity $\Delta g_{\min }$ defined by
\begin{equation}
\Delta g_{\min }=g_{\min }^{(N)}-g_{\min }^{RR}  \label{eq-xy-proof}
\end{equation}%
where $g_{\min }^{(RR)}$ is computed by (\ref{eq-4-2-11}).

It is obvious from the expression (\ref{eq-hr1}) for $g_{\min }^{(N)}$ and
the expression (\ref{eq-4-2-11}) for $g_{\min }^{RR}$, that $\Delta g_{\min
} $ defined by (\ref{eq-xy-proof}) is function of $M_{0}$ and $\theta _{w}$,
and can be computed exactly since the parameters ${{\beta _{01},}\beta
_{_{12}}^{RR}}$ and ${\beta _{_{12}}^{MR}}$ involved in (\ref{eq-hr1}) and (%
\ref{eq-4-2-11}) can be computed exactly using the two-shock and three-shock
theories.

\begin{figure} 
\begin{center}
\  \includegraphics[width=0.5\textwidth]{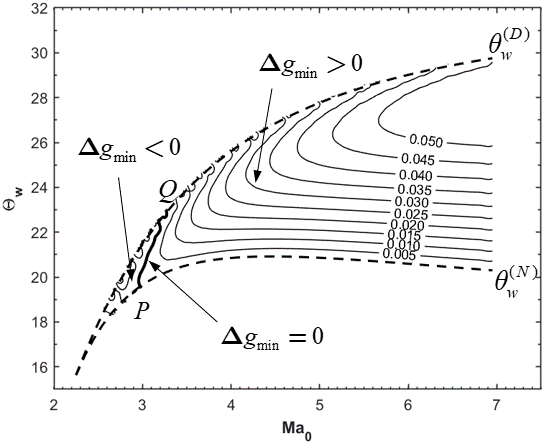}
\end{center}
\caption{Contourlines of $\Delta g_{\min }=g_{\min }^{(N)}-g_{\min }^{(RR)}$
in the dual soution domain.}
\label{fig4-2-ab}
\end{figure}

The quantity $\Delta g_{\min }$ thus obtained for $M_{0}$ and $\theta _{w}$
lying inside the dual solution domain is shown in Figure \ref{fig4-2-ab}. We
see that the inequality $\Delta g_{\min }>0$ holds for the entire region of
the dual solution domain on the right hand side of the line AB (the Mach
number at point B is around $3.5$). The region (on the left of the line AB)
with $\Delta g_{\min }<0$ is narrow.

Thus, we have proven that $g_{\min }^{(MR)}>g_{\min }^{(RR)}$ holds in the
region on the right hand side of AB. But this does not mean that $g_{\min
}^{(MR)}>g_{\min }^{(RR)}$ does not hold for the rest of the dual solution
domain. Here we provide a short proof that $g_{\min }^{(MR)}>g_{\min
}^{(RR)} $ \ holds for the entire region of the dual solution domain.

The Mach stem height $H_{T}$ is surely greater than the minimum possible
height $H_{TN}$. The mass flow across the additional height $\varepsilon $
at point O must be equal to the mass flow across the entire Mach stem, thus $%
\rho _{0}u_{0}\left( H_{TN}+\varepsilon \right) =\rho _{O}u_{O}\varepsilon $%
, which can be solved to give
\[
\varepsilon =\frac{1}{\frac{\rho _{O}u_{O}}{\rho _{0}u_{0}}-1}H_{TN}\text{.}
\]%
Here $\rho _{O}u_{O}$ is the mass flow rate at O. For subsonic flow, as is
the case here, it can be verified that
\begin{equation}
\rho _{O}u_{O}<\rho ^{\ast }u^{\ast }\text{ for any }M_{O}<1
\label{eq-xy-small}
\end{equation}%
where the $\rho ^{\ast }u^{\ast }$ denotes the critical values of $\rho u$.
Let $\lambda ^{\ast }=\frac{\rho ^{\ast }u^{\ast }}{\rho _{0}u_{0}}$, then,
due to (\ref{eq-xy-small}), we have
\begin{equation}
\varepsilon >\frac{1}{\lambda ^{\ast }-1}H_{TN},  \label{eq-xy-big}
\end{equation}%
From the isentropic flow theory for subsonic and the expression of a normal
shock wave \ (to approximate the Mach stem), it can be shown that%
\[
\lambda ^{\ast }=\lambda ^{\ast }(M_{0})=\frac{\left( 1+bc\right) ^{a}}{%
\sqrt{b}\left( 1+c\right) ^{a}}
\]%
where $a=\frac{\gamma +1}{2(\left( \gamma -1\right) },b=\frac{1+\frac{\gamma
-1}{2}M_{0}^{2}}{\gamma M_{0}^{2}-\frac{\gamma -1}{2}}$, and $c=\frac{\gamma
-1}{2}$. For $\gamma =1.4$, we have $\lambda ^{\ast }(2.3)=1.279$, $\lambda
^{\ast }(3.5)=1.446$ and the limiting value for very large inflow Mach
number is $\lambda ^{\ast }=1.66$.

\begin{figure} 
\begin{center}
\  \subfigure a) \  \includegraphics[width=0.38\textwidth]{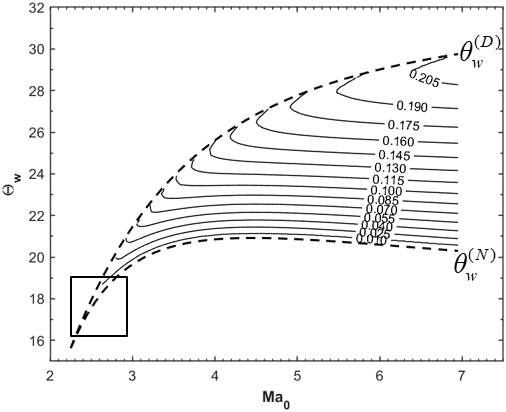} %
\subfigure b) \  \includegraphics[width=0.38\textwidth]{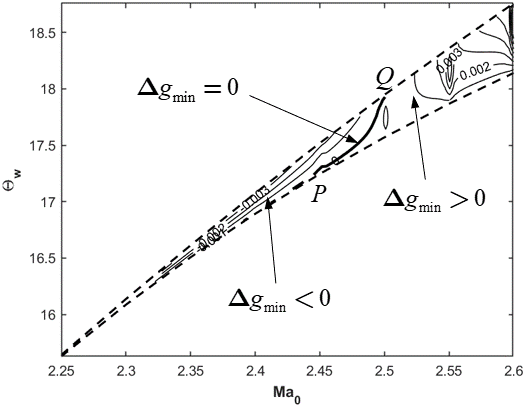}
\end{center}
\caption{Contourlines of $\Delta g_{\min }=g_{\min }^{MR}-g_{\min }^{RR}$ in
the dual soution domain£¬with $\protect \varepsilon =1.515H_{TN}.$a)
global view. b) enlarged view.}
\label{fig4-2-cd}
\end{figure}

Taking this limiting value, the inequality (\ref{eq-xy-big}) leads to
\[
\varepsilon >\frac{1}{\lambda ^{\ast }-1}H_{TN}>1.515H_{TN}
\]%
We compute $g_{\min }^{MR}$ by (\ref{eq-6-1}) with $\varepsilon =1.515H_{TN}$%
, and then $\Delta g_{\min }$ by%
\[
\Delta g_{\min }=g_{\min }^{MR}({\varepsilon =}1.515H_{TN})-g_{\min }^{RR}
\]%
The recomputed $\Delta g_{\min }$ is displayed in Figure \ref{fig4-2-cd}. We
see that $\Delta g_{\min }>0$ in the entire double solution domain.
The above short proof has assumed the Mach stem to be a normal shock wave.
This introduces a small error in the estimation of $\varepsilon $ compared
to the real Mach stem. It can be shown that this error is small enough so
the value $\varepsilon =1.515H_{TN}$ is only slightly modified, and does not
change the conclusion that $\Delta g_{\min }>0$.

\subsection{Subcritical and supercritical geometric thresholds and dual
solution stability gap}

We have proven that the inequality $\Delta g_{\min }>0$ or $g_{\min
}^{(MR)}>g_{\min }^{(RR)}$ holds within the dual solution domain. The
inequality $g_{\min }^{(MR)}>g_{\min }^{(RR)}$ means that, for the range of $%
g$ satisfying $g_{\min }^{(MR)}>g>g_{\min }^{(RR)}$, the regular reflection
is stable while the Mach reflection is unstable within the dual solution
domain. To be more explicit, there exists a range of $g$ within which the
reflected shock of regular reflection will not graze the trailing edge while
the reflected shock of Mach reflection will graze the trailing edge until it
becomes unstable.

For the above reason, we will call $g=g_{\min }^{(RR)}$ the \emph{%
subcritical geometric threshold} and $g=g_{\min }^{(MR)}$ the \emph{%
supercritical geometric threshold}, and the lower limit difference
\begin{equation}
\triangle g_{\min }^{(MR-RR)}=g_{\min }^{(MR)}-g_{\min }^{(RR)}
\label{eq-4-1-xy1}
\end{equation}%
is defined as the \emph{dual solution stability gap}.

\section{Magnitude of the stability gap}

The purpose of this section is to have an idea about how large the dual
solution stability gap may be. Due to the lack of accurate Mach stem model,
accurate estimation of this gap is not easy. We display in this section one
possible road map to find the magnitude of the stability gap, and combine
both theory and numerical simulation to obtain this quantity for one
particular condition. This quantity is further used to assess the accuracy
of dual solution stability gap based an approximate Mach stem height model.

\subsection{A road map to find $g_{\min }^{(MR)}$ and the dual solution
stability gap}

The magnitude of the stability gaps depends on $g_{\min }^{(MR)}$. Using the
expression (\ref{eq-xy-main-1}), we get the following exact expression for $%
g_{\min }^{(MR)}$,

\begin{equation}
g_{\min }^{(MR)}=f\left( M_{0},\theta _{w},g_{\min }^{(MR)}\right) +\frac{%
\phi }{\sin \theta _{w}}\left( 1-g_{\min }^{(MR)}\right)
\label{eq-xy-main-2}
\end{equation}%
where $\phi $ is defined by (\ref{xy-2-1-3}).

The expression (\ref{eq-xy-main-2}) can give exact values of $g_{\min
}^{(MR)}$ if $H_{T}=fH_{A}$ can be exactly evaluated.\ However, as mentioned
in Section 2.3, actually there is no Mach stem height model that permits us
to have a quantitative evaluation of the stability gap accurate enough for
the purpose of this paper.

Some studies have suggested that the Mach stem height is linear with $g$.
For instance, Schotz et al (1997, eq 21) obtained an approximate expression
like $\frac{H_{T}}{L}=k_{1}\frac{H_{A}}{L}+k_{2}$ where $L=w\cos \theta _{w}$%
. Bai \& Wu (2021) used numerical simulation and showed that for their
conditions $f$ can be fitted by linear expression%
\begin{equation}
f\left( M_{0},\theta _{w},g\right) =Ag+B  \label{xy-2-2-2}
\end{equation}%
with high accuracy. In (\ref{xy-2-2-2}), $A$ and $B$ depends on $M_{0}$ and $%
\theta _{w}$. \

If (\ref{xy-2-2-2}) is applied, the expression (\ref{eq-xy-main-2}) can be
written as%
\begin{equation}
g_{\min }^{(MR)}=\frac{\frac{\phi }{\sin \theta _{w}}+B}{1-A+\frac{\phi }{%
\sin \theta _{w}}}.  \label{xy-2-2-3}
\end{equation}%
Thus, if for a particular condition ($M_{0},\theta _{w}$) we have the exact
values of $A$, and $B$, then (\ref{xy-2-2-3}) can lead to exact values of $%
g_{\min }^{(MR)}$, so we can have exact values of $\triangle g_{\min
}^{(MR-RR)}$ since $g_{\min }^{(RR)}$ computed by (\ref{eq-4-2-11}) is exact.

\subsection{Magnitude of the stability gap for one particular condition}

Bai \& Wu (2021) provided fitted values of $A$ and $B$ using high fidelity
CFD, for three cases (see Table \ref{table-1}). However, it can be verified
only case 1 lies within the dual solution domain. This case ($M_{0}=4$, $%
\theta _{w}=25^{o}$) will be used to \ evaluate the magnitude of $g_{\min
}^{(MR)}$ and the dual solution stability gap.

\begin{table} 
\centerline {
\begin{tabular}{ccc}
Case & $M_{0},\theta _{w}$ & $A,B$    \\
\hline
1 & $4,25$  & $-1.031,0.723$  \\
2 & $4,30$  & $-2.370,1.699$ \\
3 & $5,25$  & $-1.854,1.431$ \\

\end{tabular}
}
\caption{Three cases with fitted values of $A$ and $B$ (Bai \& Wu 2021).}
\label{table-1}
\end{table}

Using (\ref{eq-4-2-11}), we get, for $M_{0}=4$, $\theta _{w}=25^{o}$,
\begin{equation}
g_{\min }^{(RR)}=0.239  \label{eq-3-end-1}
\end{equation}%
With $A=-1.031$ and $B=0.732$, for $M_{0}=4$, $\theta _{w}=25^{o}$, we get
from (\ref{xy-2-2-3}) the semi-theoretical value
\begin{equation}
g_{\min }^{(MR)}=0.417  \label{eq-3-end-2}
\end{equation}%
Hence, for case 1,
\[
\triangle g_{\min }^{(MR-RR)}=0.178
\]%
This means that the dual solution stability gap is as large as $74\%$ of $%
g_{\min }^{(RR)}$, so the supercritical geometric threshhold is largely
separated from the subcritical one, i.e., there is a risk of dynamic
transition over a wide range of $g$ for case 1.

\begin{figure} 
\centering a)\subfigure{\includegraphics[width=0.5\textwidth]{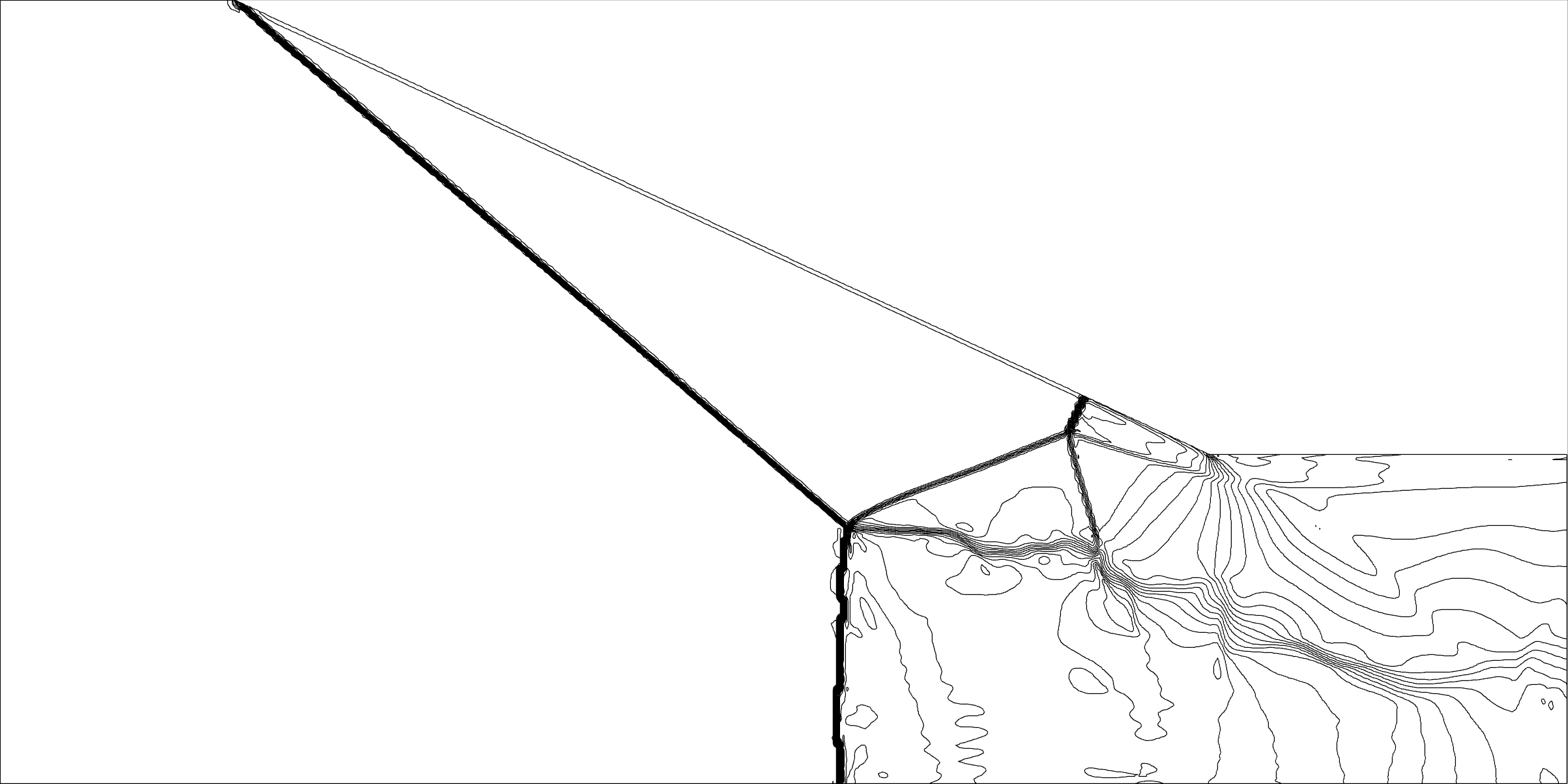}} b)%
\subfigure{\includegraphics[width=0.5\textwidth]{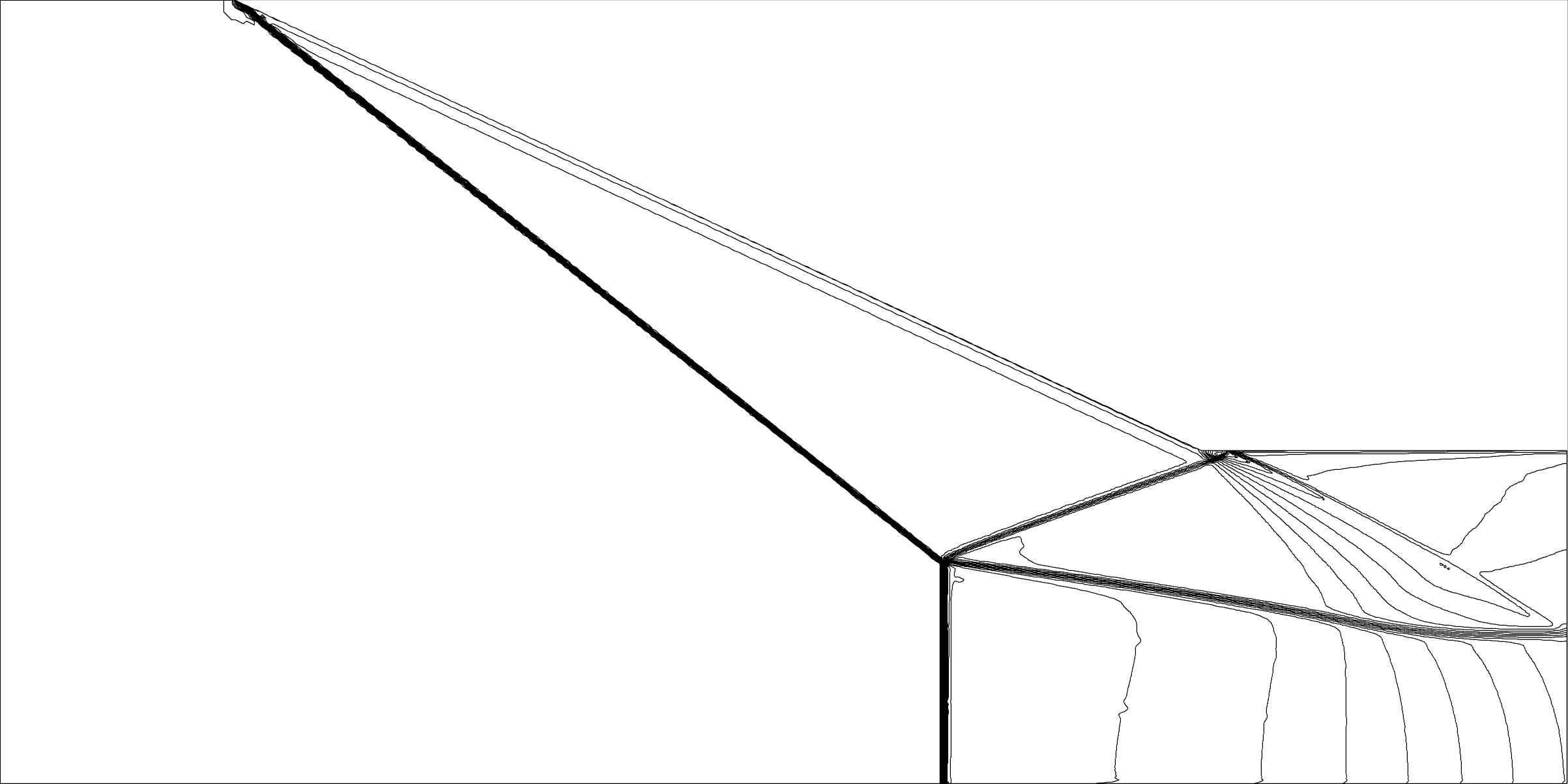}}
\caption{Mach number contours for Case 1: a) $g=0.420$, b) $g=0.425$.}
\label{fig-3-1}
\end{figure}

Now we further check whether the magnitude $g_{\min }^{(MR)}=0.417$ is
accurate enough. For this purpose, we perform numerical simulation for a
series of values of $g$. The Mach number contour lines for two values of $g$
around the theoretical value $g_{\min }^{(MR)}=0.417$ are given in Figure %
\ref{fig-3-1}. It is seen that for $g=0.42$, we get unstable results, as can
be seen from Figure \ref{fig-3-1} (a) which gives a snapshot of the time
dependent results. This means that the lower limit $g_{\min }$ will be
higher than $0.42$. According to Figure \ref{fig-3-1} (b), the reflected
shock will not graze the trailing edge for $g=0.425$, which means that $%
g_{\min }$ will be lower than $0.425$. Thus, $0.42<g_{\min }<0.425$. Thus, $%
g_{\min }^{(MR)}=0.4225\pm 0.0025$. This is $(1.\, \allowbreak 32\pm 0.60)$ $%
\%$ larger than the semi-theoretical value given by (\ref{eq-3-end-2}). This
agreement is well enough, considering the possible error in the estimation
of $A$ and $B$ and in numerical simulation.

\begin{table} 
\centerline {
\begin{tabular}{llc}
Case & $M_{0},\theta _{w}$ &$g$  \\
\hline
1 & $4,25$   & $0.420, 0.425, 0.475$ \\
2 & $4,30$   & $0.500,0.550,0.600$ \\
3 & $4,25$   & $0.550,0.575,0.600$ \\

\end{tabular}
}
\caption{The values of $g$ for each set of conditions, used in CFD.}
\label{table-2}
\end{table}

\begin{figure} 
\begin{center}
\  \subfigure a) \  \includegraphics[width=0.38\textwidth]{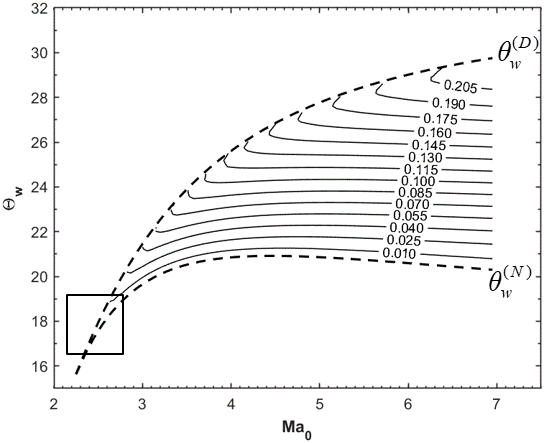} %
\subfigure b) \  \includegraphics[width=0.38\textwidth]{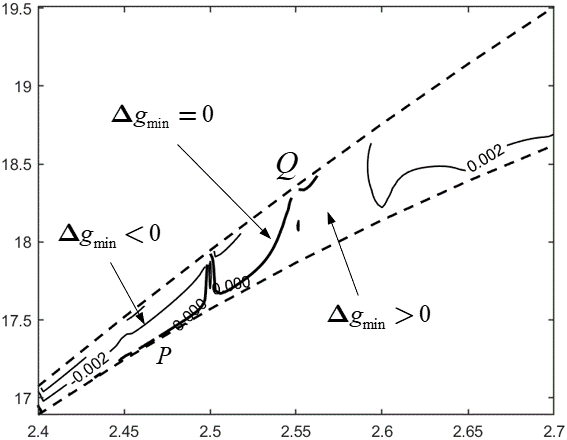}
\end{center}
\caption{Contourlines of  $\triangle g_{\min }^{(MR-RR)}$ in the $M_{0}-%
\protect \theta _{w}$ plane. The Mach stem height is provided by the model of
Bai \& Wu (2021).  a) global view. b) enlarged view.}
\label{fig4-1-2}
\end{figure}

We have been able to give quantitatative information about the magnitude of
dual solution stability gap for only one condition. One may expect to have
all quatitative information within the entire dual solution domain. However,
as stated previously, this is actually impossible since we lack a Mach stem
height model accurate enough for this purpose. To see the large errors due
to the use of an approximate Mach stem height model, we use the model of Bai
\& Wu (2021, Eq. (25)) which provided explicit expressions for $A$ and $B$.
With this approximative model, we obtain $\triangle g_{\min }^{(MR-RR)}$ as
displayed in Figure \ref{fig4-1-2}. We see that the difference $\triangle
g_{\min }^{(MR-RR)}$ thus estimated is larger than that based on the minimal
Mach stem height (see Figure \ref{fig4-2-cd}). Now consider case 1 of \ref%
{table-1}, for which the confirmed accurate value is $\triangle g_{\min
}^{(MR-RR)}=0.178$, while $\triangle g_{\min }^{(MR-RR)}\approx 0.12$ based
on the approximate Mach stem height model. The latter is about $33\%$
smaller than the accurate value, meaning that the dual solution stability
gap based on an approximate Mach stem height model could have unacceptable
errors.

The road map of finding $g_{\min }^{(MR)}$ and the dual solution stability
gap discussed in section 3.1 is actually a means that compromises accuracy
and efficiency. Under the assmption that the relative Mach stem height is
linear with $g$, only two computations are required to determine $A$ and $B$
for each condition ($M_{0},\theta _{w}$), and then (\ref{xy-2-2-3}) and (\ref%
{eq-4-2-11}) are used to compute $g_{\min }^{(MR)}$ and $g_{\min }^{(RR)} $,
necessary to compute $\triangle g_{\min }^{(MR-RR)}$. This is not only
accurate enough, but also much more efficient than simply using numerical
simulation (which requires far more than two computations with different $g$
to find $g_{\min }^{(MR)}$).

\section{Dynamic transition from stable regular reflection to unstable Mach
reflection within the dual solution stability gap}

The objective of this section is to demonstrate a new type of dynamic
transition within the dual solution stability gap, using numerical
simulation, and to display possible shock reflection or interaction patterns
during the process of dynamic transition.

\subsection{Dynamical transition problem}

We have seen from section 4.1 that the supercritical geometric threshold $%
g_{\min }^{(MR)}$ is larger than the subcritical one $g_{\min }^{(RR)}$,
i.e., there exists a dual solution stability gap $\triangle g_{\min
}^{(MR-RR)}=g_{\min }^{(MR)}-g_{\min }^{(RR)}>0$ away from $\theta
_{w}=\theta _{w}^{(N)}$. This means that, at any point $\theta
_{w}^{(N)}(M_{0})<\theta _{w}<\theta _{w}^{(D)}(M_{0})$, if the relative
trailing edge height $g$ satisfies%
\begin{equation}
g_{\min }^{(RR)}<g<g_{\min }^{(MR)},  \label{eq-4-2}
\end{equation}%
then regular reflection solution is stable, and Mach reflection solution is
unstable. This would mean a new dynamic transition possibility: stable
regular reflection may transit to unstable Mach reflection when there is
large amplitude disturbance.

Before this is made clearer, we recall what happens in conventional dynamic
transition, for which stable regular reflection transits to stable Mach
reflection by large amplitude disturbance, for $g$ outside the dual solution
stability gap.

It is known that, in the dual solution domain, RR may transit to MR if the
amplitude of the disturbance exceeds a certain level (Ivanov et al. 1997,
1998, 2000). The dynamic transition process can be understood by following
the time history of this transition using numerical simulation. Various
forms of disturbance have been considered. Ivanov et al. (1997) considered
disturbances in the form of strong short-time changes in the free-stream
velocity. Kudyavtev et al. (2002) used three types of upstream perturbation:
inlet pressure wave disturbances (including shock waves and rarefaction
waves), inlet contact discontinuity disturbances, and localized density
disturbances. All these types of disturbances have been shown to be able to
cause dynamic transition from regular reflection to Mach reflection.

The dynamic transition process has been also studied theoretically. Mouton
\& Hornung (2007) assumed a single but evolutionary MR, which satisfies a
steady flow model when the reference frame is attached to the moving triple
point and built a dynamical transition model that tracks the growth of Mach
stem height during the transition. Li, Gao \& Wu (2011) followed the work of
Mouton \& Hornung (2007) and included more fine structures than a single
unsteady MR to build a dynamic transition model. They also used a local
discontinuity to force the transition from regular reflection to Mach
reflection, and the initial-period Riemann solution of the local
discontinuity interacts with the initial regular reflection to evolve the
flow to steady Mach reflection.

\subsection{Dual solution unstable condition and disturbance to inaugurate
dynamic transition}

Here we consider dynamic transition in the dual solution domain, for
geometric setup which has a trailing edge height above the lower limit $%
H_{R,\min }$ for regular reflection, but below the lower limit $H_{R,\min }$
for Mach reflection. In this situation, RR should transit to unstable MR,
unlike dynamic transition previously studied, in which RR transit to stable
MR.

For numerical simulation, we choose the specific condition $M_{0}=4,\theta
_{w}=25^{o}$. We have shown, in section 4, that
\[
g_{\min }^{(RR)}=0.239\text{, }g_{\min }^{(MR)}=0.417
\]%
for this condition. Thus, if we choose a $g$ such that $0.239<g<0.417$, then
under some disturbance, regular reflection could transit to unstable Mach
reflection, in contrast to transition from regular reflection to stable Mach
reflection studied earlier for which $g$ is above the lower limit. The value
$g=0.328$ meets such a condition, i.e., with $g=0.328$, regular reflection
could transit to unstable Mach reflection.

Following Kudyavtev et al. (2002), we use a contact discontinuity
disturbance in the inlet. Starting from the steady numerical solution of
regular reflection, we put in the inlet for a time interval $t_{disturb}$, a
disturbance of the density $\bigtriangleup \rho /\rho _{0}$. Since this is a
contact discontinuity, the pressure and velocity are kept unchanged. This
disturbance then causes a disturbance of the Mach number $M^{\prime }=M_{0}%
\sqrt{1+\bigtriangleup \rho /\rho }$ so the disturbance $\bigtriangleup \rho
<0$ induces a decrease of Mach number to be above the detachment condition
so that transition from regular reflection to Mach reflection is made
possible. \ The disturbance is either given in the entire height of the
inlet, or is localized by giving the disturbance only in a height of $\frac{1%
}{20}H_{A}$ counting from the reflecting surface (about 10 cells). The
duration $t_{disturb}$ of disturbance, measured with $\tau _{disturb}=\frac{%
t_{disturb}}{H_{A}/a_{0}}$ where $a_{0}$ is the sound speed at the inlet, is
also a factor to be considered.

Several test cases with various values of these factors are given in Table %
\ref{table-4-1}. Numerical simulation shows three situations: failure to
inaugurate any Mach reflection, transition from RR to MR to unstart flow
(here called direct transition), and transition from RR to MR to RR (here
called inverted transition).

\begin{table} 
\centerline {
\begin{tabular}{clrrrl}
Case & disturbance&  $\Delta \rho/\rho_{0}$& $M'$ & $\tau_{disturb}$ & Transition mode \\
\hline
1 & localized  & $-0.250$  & $3.464$ & $0.069$ & Failed (with RR $\rightarrow$ perturbed RR $\rightarrow$ RR)\\
2 & localized  & $-0.375$  & $3.162$ & $0.069$ & Sucess (with RR $\rightarrow$ MR $\rightarrow$ unstart)\\
3 & localized  & $-0.500$  & $2.828$ & $0.069$ & Succes (with RR $\rightarrow$ MR $\rightarrow$ unstart)\\
4 & localized  & $-0.250$  & $3.464$ & $0.121$ & Succes (with RR $\rightarrow$ MR $\rightarrow$ unstart)\\
5 & localized  & $-0.250$  & $3.464$ & $0.174$ & Succes (with RR $\rightarrow$ MR $\rightarrow$ unstart)\\
6 & localized  & $-0.438$  & $3.000$ & $0.296$ & Succes (with RR $\rightarrow$ MR+type IV SI  $\rightarrow$ unstart)\\
7 & full inlet  & $-0.438$  & $3.000$ & $0.174$ & Failed (with RR $\rightarrow$ MR $\rightarrow$ RR) \\
\end{tabular}
}
\caption{Dynamical transition in the double solution domain with \ $(M_{0},%
\protect \theta _{w})=(4$, $25^{o})$ (at which $g_{\min }^{RR}=0.239$, $%
g_{\min }^{MR}=0.417$) and $g=0.328$. Type IV SI means type IV shock
interference.}
\label{table-4-1}
\end{table}

For case 1, the disturbance is applied locally in a stripe close to the
reflected surface, with relatively small discontinuity and short duration.
We observe no transition, i.e., no Mach reflection structure is produced at
any time. For other conditions, we have either direct transition or inverted
transition.

\subsection{Direct dynamic transition type I: from RR to MR to unstart}

\begin{figure} 
\begin{center}
\subfigure a)\includegraphics[width=0.45\textwidth]{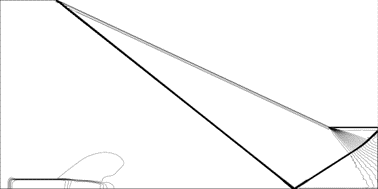} \  \subfigure %
b)\includegraphics[width=0.45\textwidth]{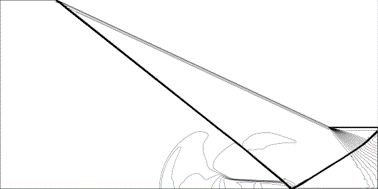} \subfigure c)%
\includegraphics[width=0.45\textwidth]{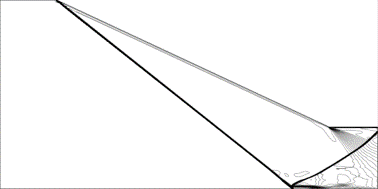} \subfigure d)%
\includegraphics[width=0.45\textwidth]{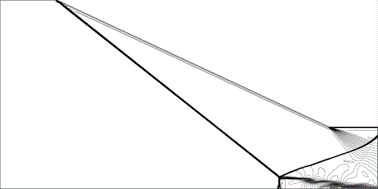} \subfigure e)%
\includegraphics[width=0.45\textwidth]{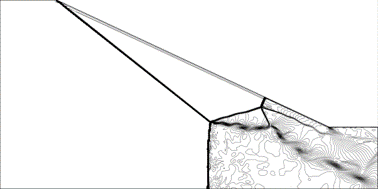} \subfigure f)%
\includegraphics[width=0.45\textwidth]{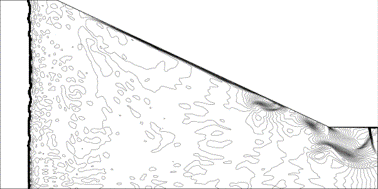}
\end{center}
\caption{Mach number contours for direct dynamic transition type I (case 4
in Table \protect \ref{table-4-1}). }
\label{fig4-1bis}
\end{figure}

Compared to case 1, Cases 2 and 3 increase the intensity of the density
disturbances. We observe "RR to MR to unstart" transition. For Case 4 and
Case 5, the duration of density is increased compared to Case 1, and we also
observe "RR to MR to unstart" transition. Figure \ref{fig4-1bis} displays
for Case 4 the Mach number at several stages of the transition process.

Stage 1: initial RR. The RR result, not displayed, is treated as the initial
condition before the upstream disturbance is introduced.

Stage 2: disturbance propagation. The upstream disturbance is generated at
the inlet, and then propagates toward the reflecting point. Figure \ref%
{fig4-1bis} (a) is the result at some instance. This upstream disturbance
has not yet touched the reflection point so the regular reflection
configuration near the reflecting point is not yet affected.

Stage 3: disturbance RR interaction stage. The disturbance reaches the
reflecting point and strengthens the incident shock at the reflecting point
(Figure \ref{fig4-1bis} (b)). Locally, the shock angle of the incident shock
overtakes the detachment condition so the local RR structure transits to MR
(Figure \ref{fig4-1bis} (c)).

Stage 4: pseudo-steady MR stage (Figure \ref{fig4-1bis} (d)). The density
disturbance has fully transmitted the Mach stem and there remains a pure
pseudo-steady MR structure. For conventional dynamic transition as
considered by Kudyavtev et al. (2002) and Mouton \& Hornung (2007), the MR
will become stable. Here, since $g$ lies within the dual solution stability
gap, the MR can not be stabilized, and it will propagates towards the
upstream direction.

Stage 5: unsteady double MR stage. The reflected shock of the pseudo-steady
MR, after grazing the trailing edge, reflects at the lower wedge surface,
and creating another pseudo-steady MR structure for the present condition.
The lower MR and upper MR both propagates toward the inlet (Figure \ref%
{fig4-1bis} (e)).

Stage 6: unstart subsonic flow (Figure \ref{fig4-1bis} (f)). The double MR
structure has touched the inlet and a shock is formed at the inlet. This
shock would become a bow shock once a steady state could be reached. The
flow downstream becomes subsonic, and corresponds to what we call unstart
flow.

\subsection{Direct dynamic transition type II: from RR to MR + type IV shock
interference to unstart}

\begin{figure} 
\begin{center} 
\subfigure a)\includegraphics[width=0.45\textwidth]{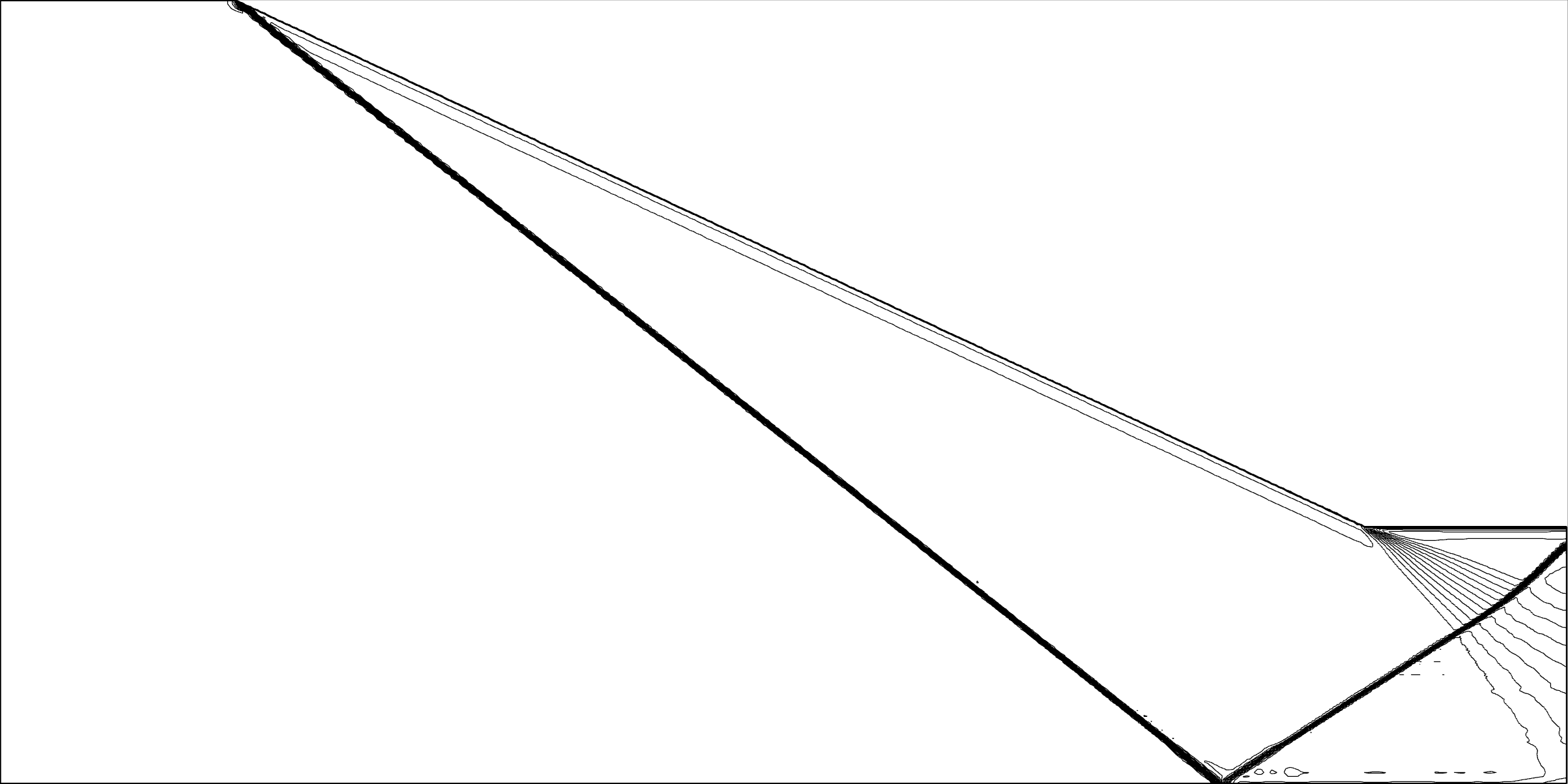} \ 
\subfigure b)\includegraphics[width=0.45\textwidth]{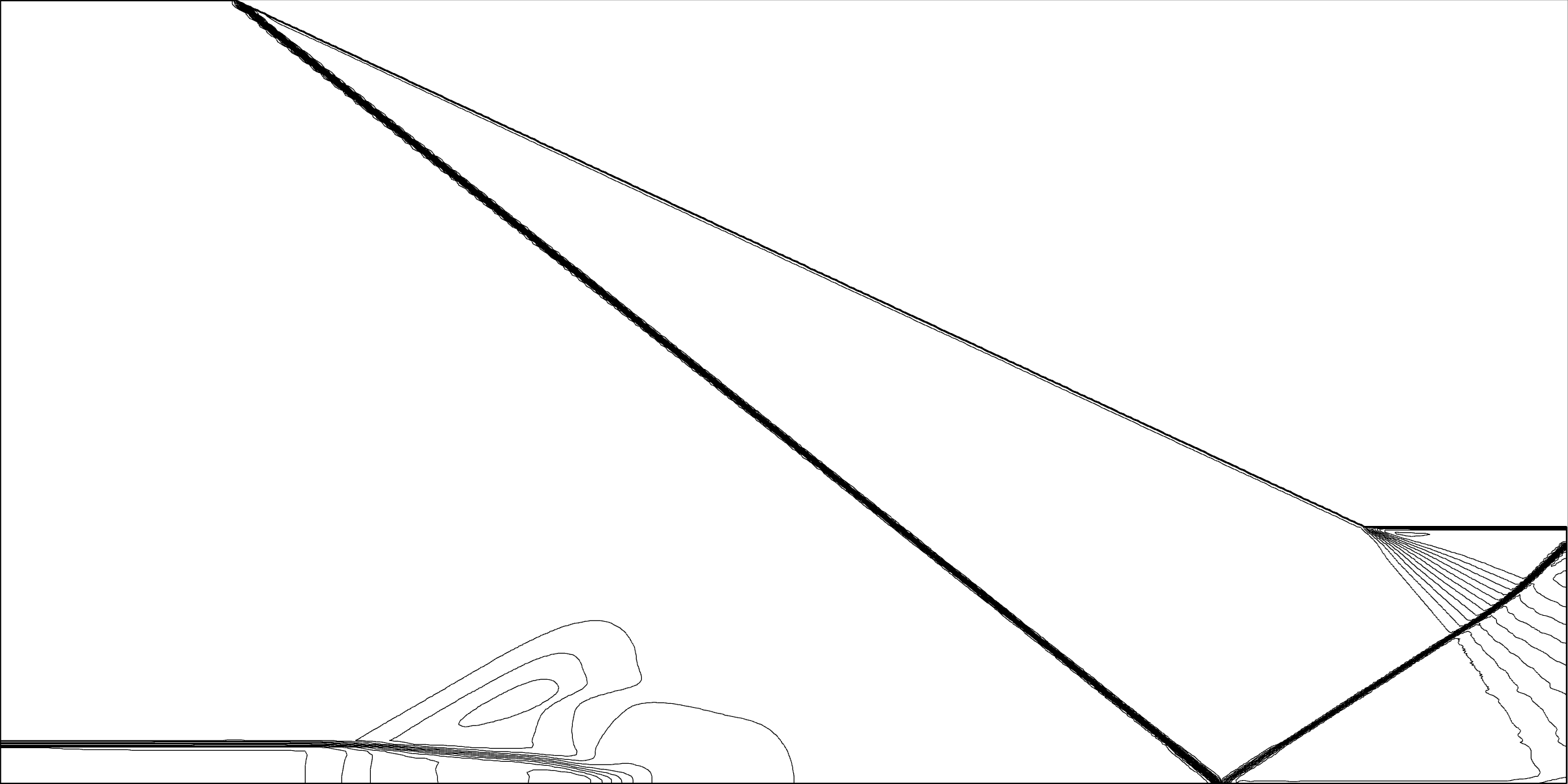} \\
\subfigure c) \includegraphics[width=0.45\textwidth]{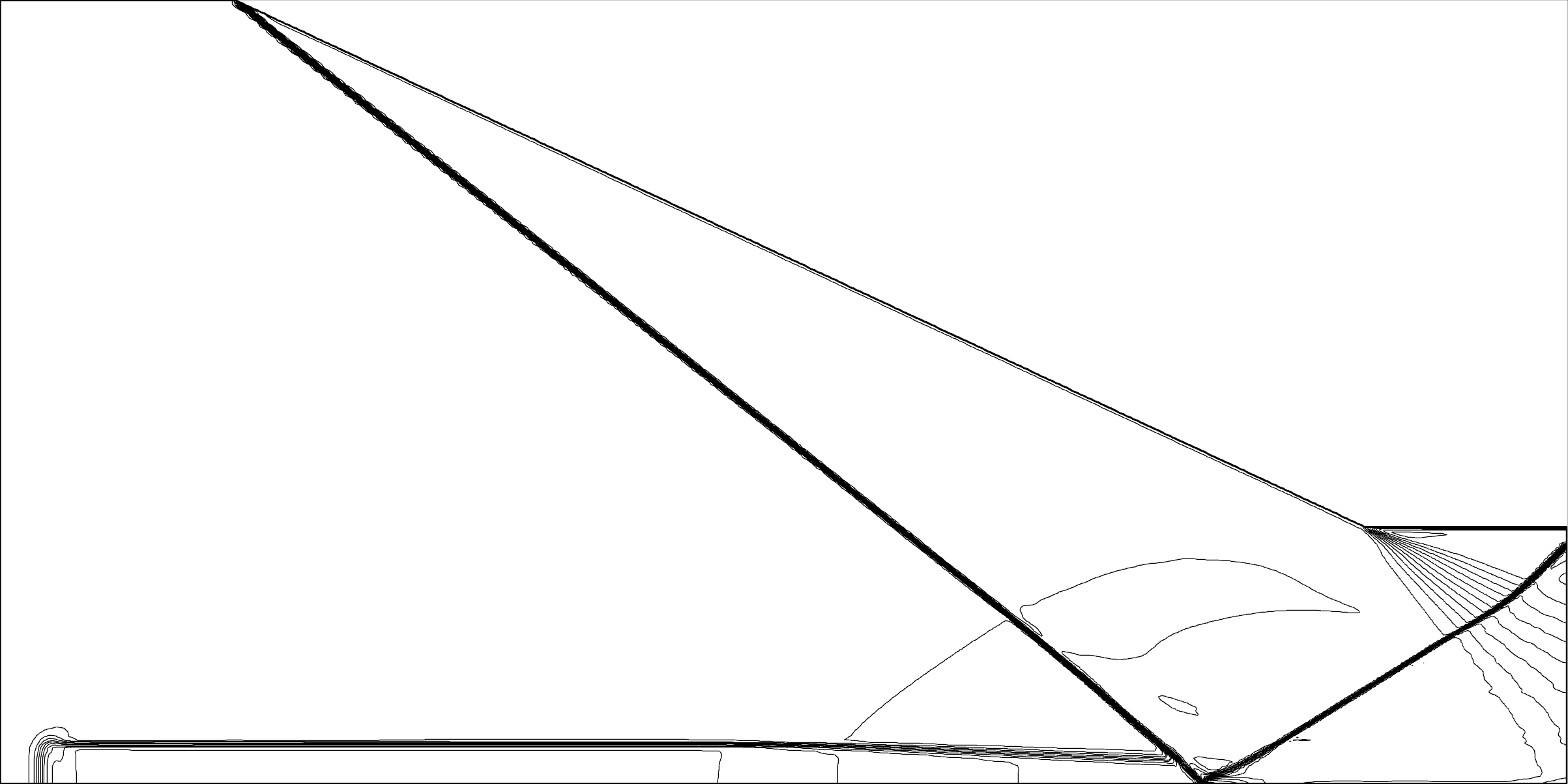}  
\subfigure d)\includegraphics[width=0.45\textwidth]{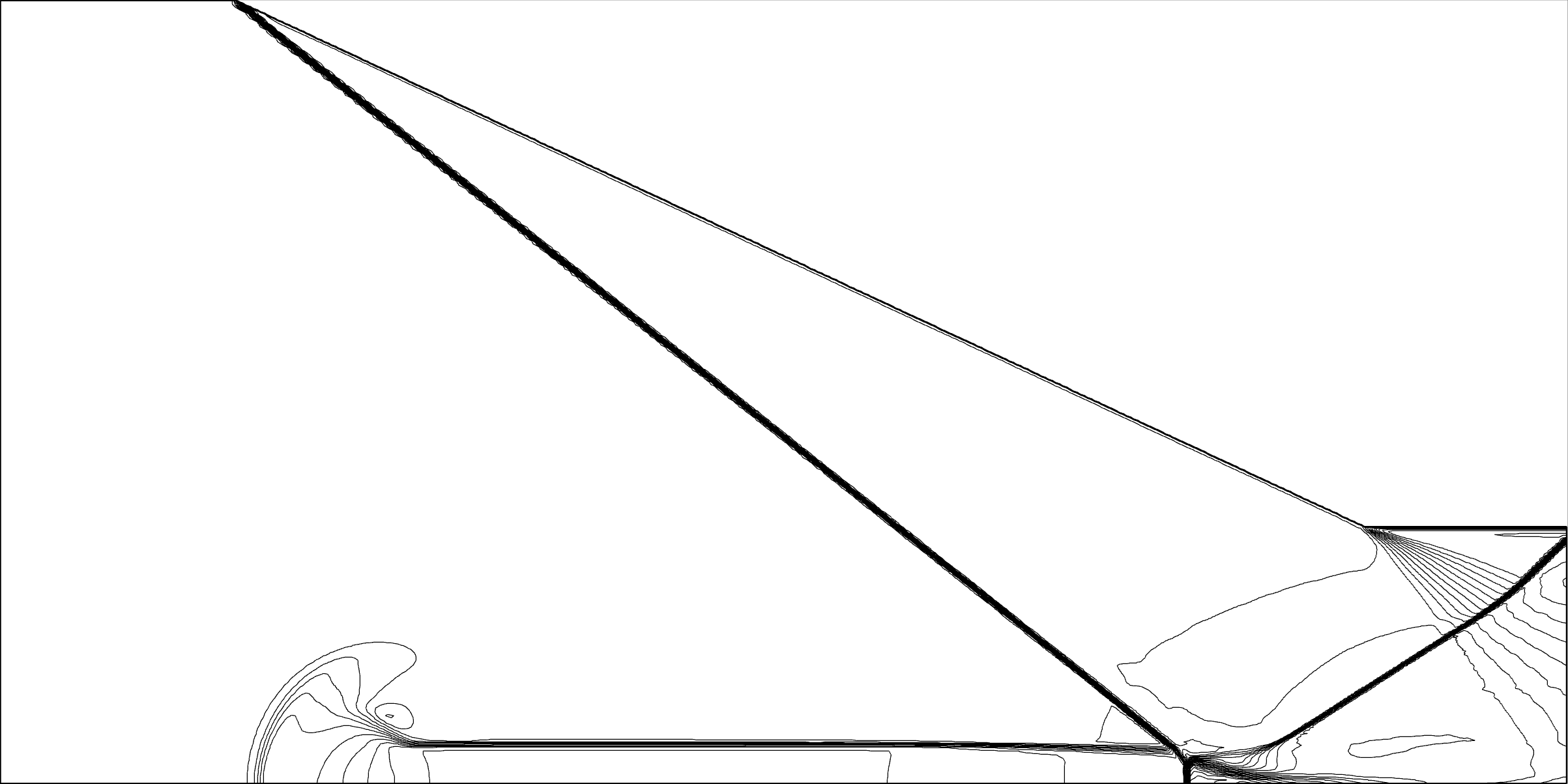} \\
\subfigure e) \includegraphics[width=0.45\textwidth]{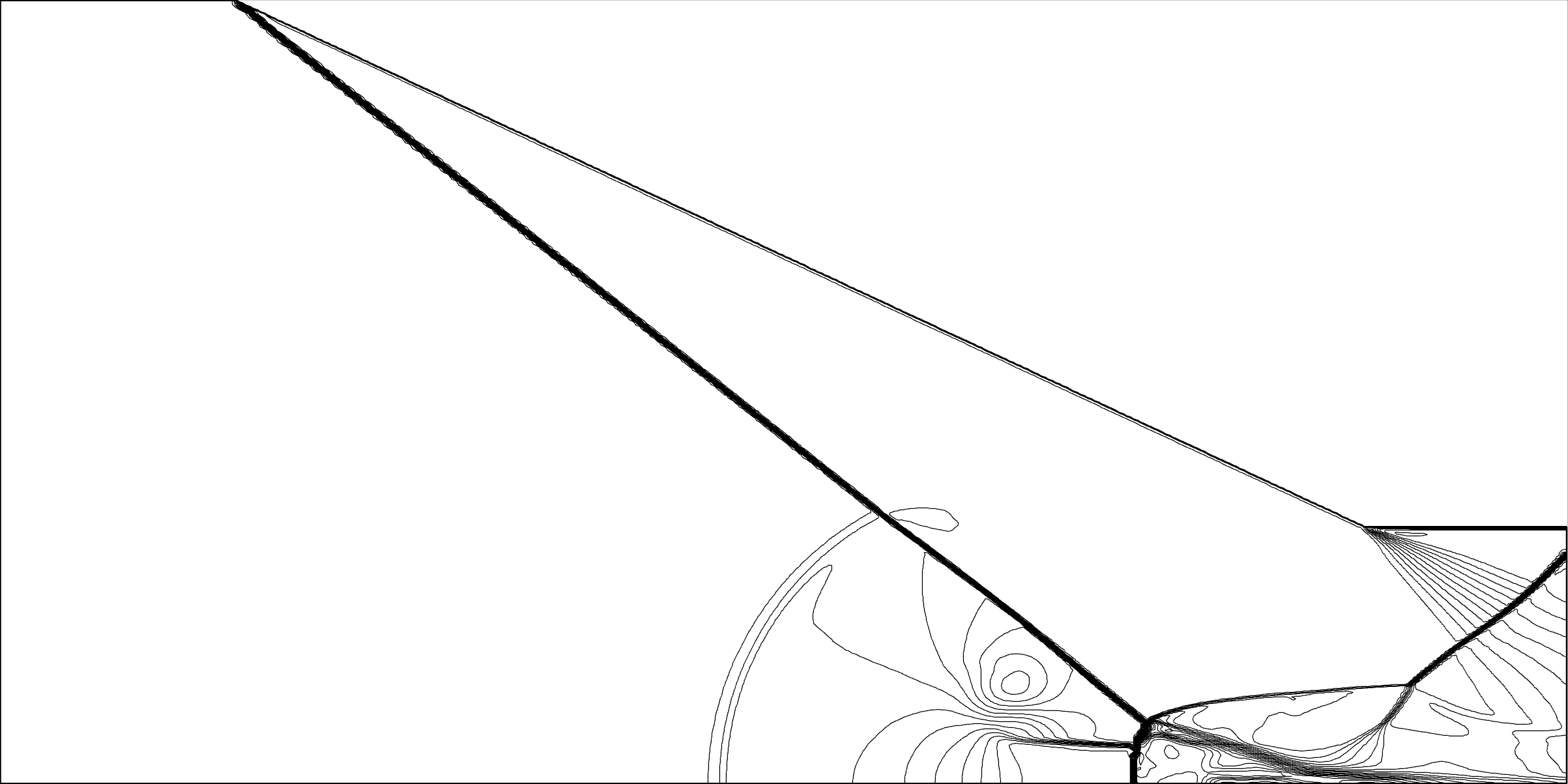} 
\subfigure f) \includegraphics[width=0.45\textwidth]{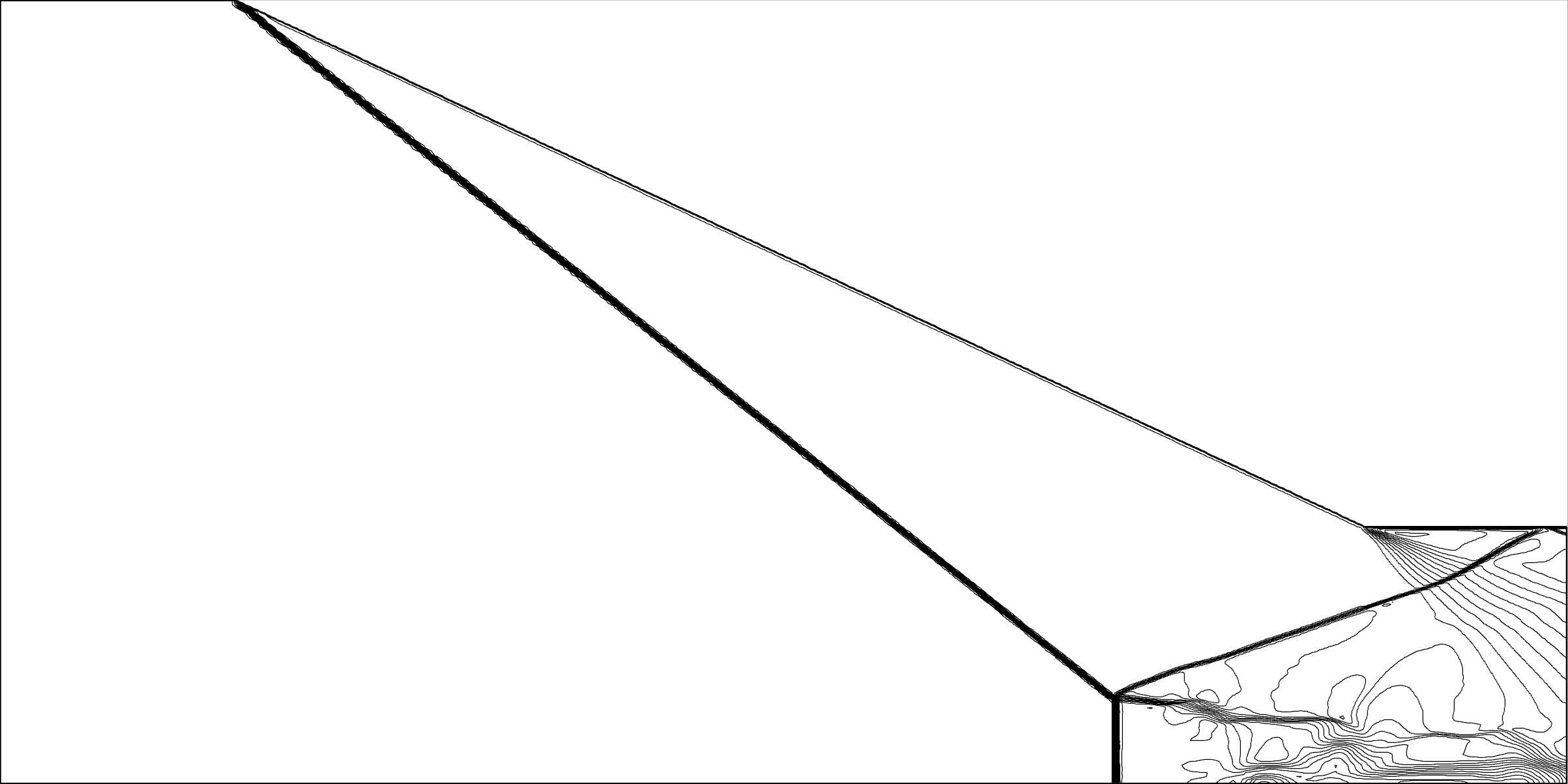} \\
\subfigure g) \includegraphics[width=0.45\textwidth]{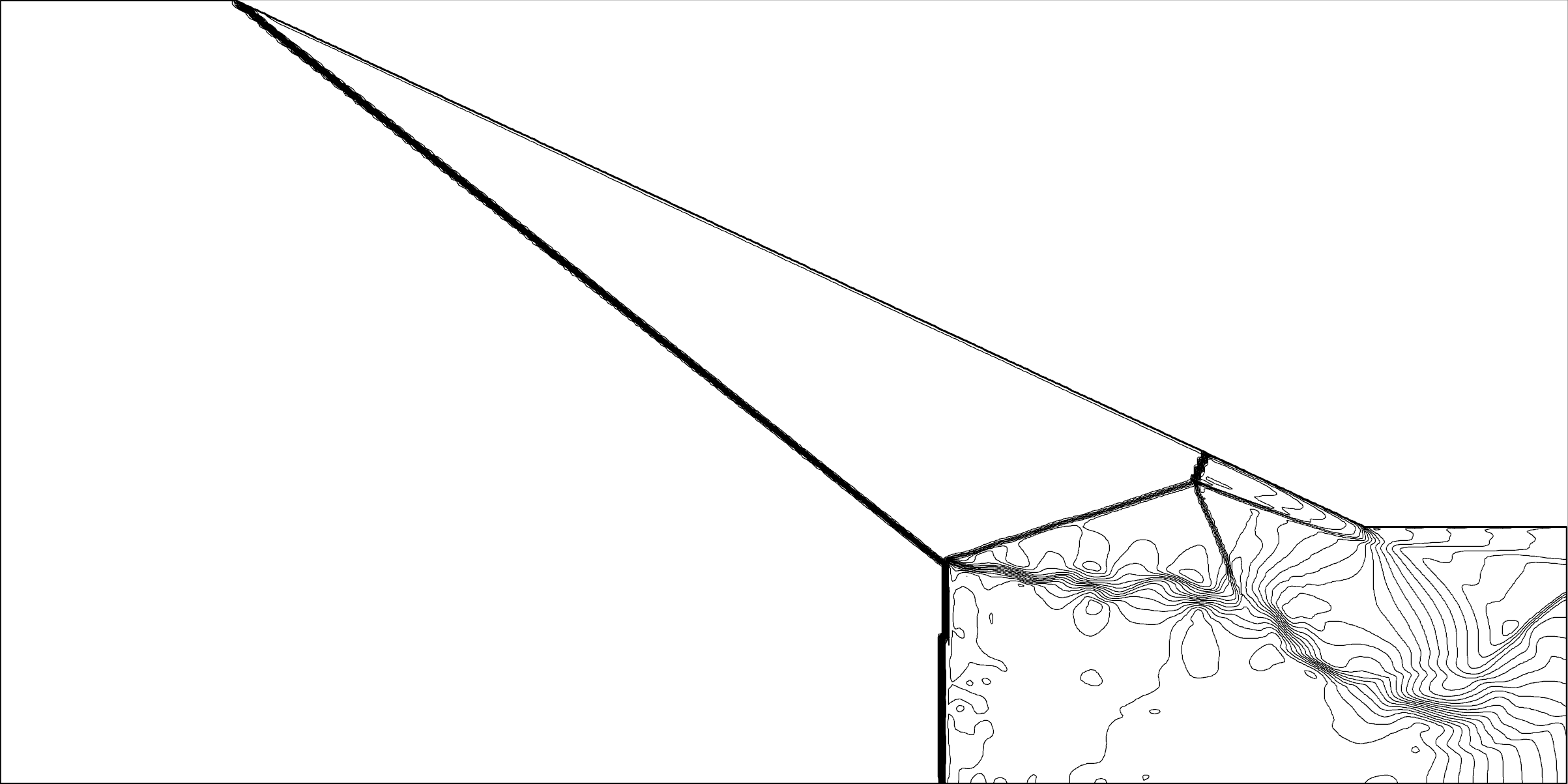} 
\subfigure h) \includegraphics[width=0.45\textwidth]{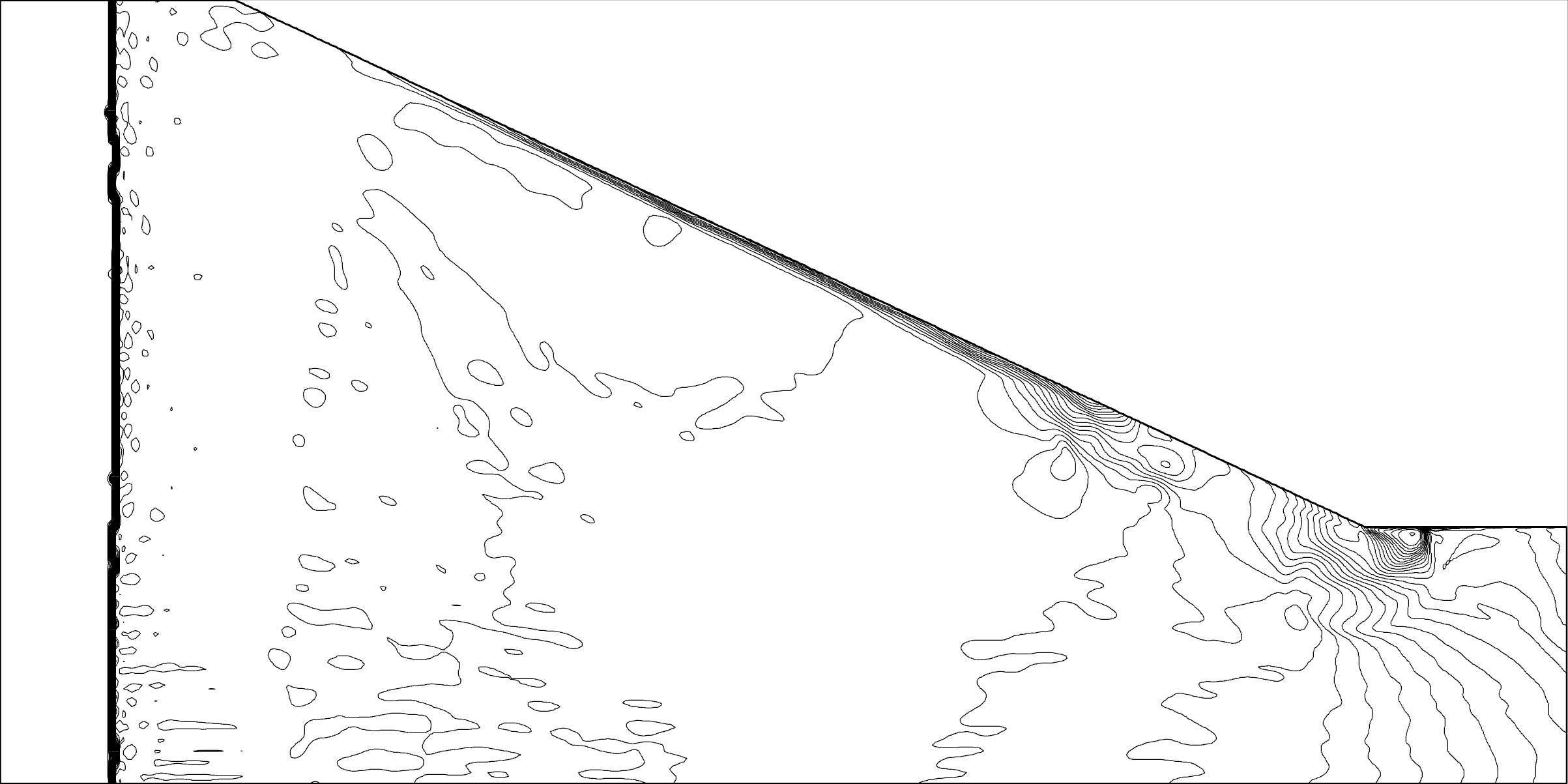} \\
\end{center} 
\caption{Mach number contours for direct dynamic transition (case 6 in Table \protect \ref{table-4-1}). } \label{fig4-2}
\end{figure}

For Case 6, both the intensity of density disturbance and the duration of
disturbance are increased compared to Case 1. We observe the so-called
direct transition type II, as displayed in Figure \ref{fig4-2}, which shows
the Mach number at several stages of the transition process.

Stage 1: initial RR. The RR result, shown in Figure \ref{fig4-2} (a), is
treated as the initial condition before the upstream disturbance is
introduced.

Stage 2: disturbance propagation. The upstream disturbance is generated at
the inlet, and then propagates toward the reflecting point. Figure \ref%
{fig4-2} (b) is the result at $t\approx \frac{1}{2}t_{disturb}$. This
upstream disturbance has not yet touched the reflection point so the regular
reflection configuration near the reflecting point is not yet affected.

Stage 3: disturbance RR interaction stage. The disturbance reaches the
reflecting point and strengthens the incident shock at the reflecting point
(Figure \ref{fig4-2} (c)). For the present condition, this occurs at $%
t\approx t_{disturb}$, when the upstream disturbance at the inlet is
terminated. Locally, the shock angle of the incident shock overtakes the
detachment condition so the local RR structure transits to MR (Figure \ref%
{fig4-2} (d)).

\begin{figure} 
\begin{center}
\includegraphics[width=0.45\textwidth]{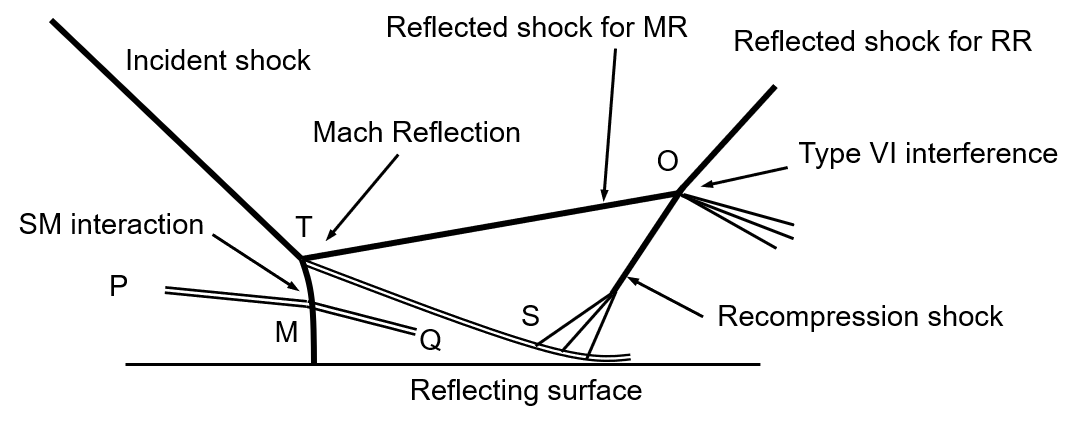}
\end{center}
\caption{Schematic display of the MR+type VI interference. SM interaction
means the interaction between the edge of the density disturbance (slipline
PM) and the Mach stem, which gives a transmitted slipline (MQ).}
\label{fig-xy-4-2}
\end{figure}

Stage 4: disturbed MR + type VI interference stage (Figure \ref{fig4-2} (e),
see Figure \ref{fig-xy-4-2} for a schematic display). Note that type VI
interference does not appear in direct dynamic transition type I. The Mach
stem of MR is still subjected to the interaction of the upstream density
disturbance. This disturbance has a slipline (PM) almost parallel to the
reflecting surface. The interaction between this slipline and the Mach stem,
at point M, leads to a transmitted slipline (MQ). The type VI interference
structure is composed of the reflected shock of MR, and a recompression
shock over the turning point S of the slipline of MR. This interference
leads to a shock that is one part of the reflected shock of the original RR.
The recompression shock is due to the flow, which is initially parallel to
the slipline, will be deflected to be parallel to the reflecting surface.
Note that type VI interference also appears in the conventional dynamic
transition problem studied by Kudyavtev et al. (2002) and Li, Gao \& Wu
(2011).

Stage 5: pseudo-steady MR. The density disturbance has fully transmitted the
Mach stem. Both this disturbance and the type VI interference structure have
propagated far downstream, so there remains a pure pseudo-steady MR
structure (Figure \ref{fig4-2} (e)). For conventional dynamic transition as
considered by Kudyavtev et al. (2002) and Mouton \& Hornung (2007), the MR
will become stable. Here, since $g$ lies within the dual solution stability
gap, the MR can not be stabilized, and it will propagates towards the
upstream direction.

Stage 6: unsteady double MR stage. The reflected shock of the pseudo-steady
MR, after grazeing the trailing edge, reflects at the lower wedge surface,
and creating another pseudo-steady MR structure for the present condition.
The lower MR and upper MR both propagate toward the inlet (Figure \ref%
{fig4-2} (f)).

Stage 7: unstart subsonic flow (Figure \ref{fig4-2} (g)). The double MR
structure has touched the inlet and a  shock is formed at the inlet. The
flow downstream becomes subsonic, and corresponds to what we call unstart
flow.

\subsection{Inverted\ transition: from RR to MR to RR}

\begin{figure} 
\begin{center}
\subfigure a) \includegraphics[width=0.45\textwidth]{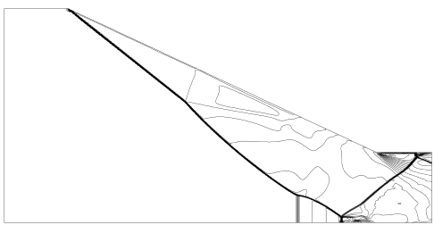} \ %
\subfigure b) \includegraphics[width=0.45\textwidth]{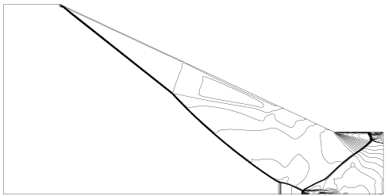} \\ %
\subfigure c) \includegraphics[width=0.45\textwidth]{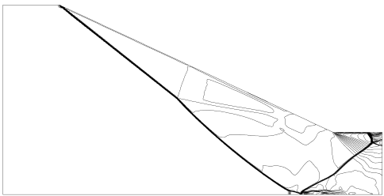} \ %
\subfigure d) \includegraphics[width=0.45\textwidth]{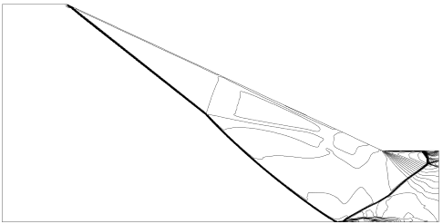} \\ %
\subfigure e) \includegraphics[width=0.45\textwidth]{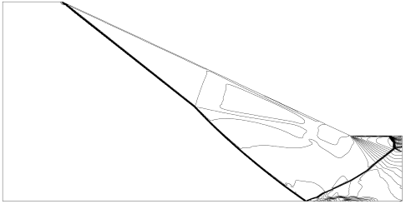} \ %
\subfigure f) \includegraphics[width=0.45\textwidth]{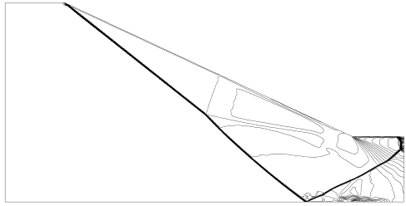} \\%
\end{center}
\caption{Mach number contours for inverted dynamic transition (case 7 in
Table \protect \ref{table-4-1}). }
\label{fig4-2-bis}
\end{figure}

Figure \ref{fig4-2-bis} shows the results for case 7. The disturbance is
applied to the whole inlet so the entire incident shock will be disturbed.
We observed what we call here inverted transition.

Stage 1: RR to MR transition. This incident shock will be strengthened so it
causes RR to MR transition once the disturbance has reached the reflecting
point. Figure \ref{fig4-2-bis}(a) displays the result at a moment when RR
has transited to MR.

Stage 2: weakening MR. The highly disturbed part of the incident shock is
weakened by the ending part of the density disturbance, so the Mach stem
reduces its height (Figure \ref{fig4-2-bis}(b)-(c)).

Stage 3: MR back into RR. The transition process is inverted, and the MR
transits back into a highly disturbed RR (Figure \ref{fig4-2-bis}(d)-(f)).

Stage 4: stable RR. Finally, we get back the initial RR structure.

The inverted dynamic transition occurs when the density disturbance first
strengthens the incident shock (so that detachment condition is reached) and
then weakens the incident shock (so that von Neumann condition is reached).

\section{Conclusions}

In this paper, we have studied the lower limit $H_{R,\min }$ of the relative
trailing edge height, at which the reflective shock grazes the trailing edge
and below which shock reflection may become unstable. We have particularly
considered $H_{R,\min }^{(RR)}$ and $H_{R,\min }^{(MR)}$, i.e., the lower
limits for RR and MR.

A major work is that we have proved that $H_{R,\min }^{(MR)}>H_{R,\min
}^{(RR)}$ holds in the entire dual solution domain. This would have not been
possible since actually any Mach stem height is not quantitatively correct.
To overcome this difficulty, we show that this holds if we can prove it
using the minimum Mach stem height. The proof shows that $H_{R,\min
}^{(MR)}>H_{R,\min }^{(RR)}$ indeed holds with the minimum Mach stem height,
which can be exactly given.

We have thus identified a dual solution stability gap for $g$ between the
lower limit $H_{R,\min }^{(RR)}$ (called subcritical threshold) and the
lower limit $H_{R,\min }^{(MR)}$ (called supcritical threshold). Above the
supercritical threshold, both RR and MR can be stable, i.e., we may have
steady stable RR and MR solution in the dual solution domain. Below the
subcritical threshold, both RR and MR are unstable.

A road map is given to obtain the quantity of the dual solution stability
gap. This relies on the use of a linear Mach stem height assumption (an
assumption verified by past studies) and numerical simulation to determine
the linear coefficients. For the particular condition ($M_{0}=4$ and $\theta
_{w}=25^{o}$) for which Bai \& Wu (2021) provided fitted data for the linear
coefficients, we get $H_{R,\min }^{(MR)}=$ $0.417H_{A}$ $\ $and $H_{R,\min
}^{(RR)}$ $=0.239H_{A}$, which means that the dual solution stability gap is
large compared to the subcritical geometric threshold $H_{R,\min }^{(RR)}$.
This quantity is further confirmed by numerical simulation, and is used to
show that the road map compromises the accuracy and efficiency to determine
the dual solution stability gap (pure theoretical estimation based on an
approximate Mach stem height model has large errors, while pure numerical
simulation requires a large amount of $H_{R}$ to find the stability gap).

Within the dual solution stability gap, i.e., for $H_{R,\min
}^{(RR)}<H_{R}<H_{R,\min }^{(MR)}$, RR is stable and MR is unstable. This
means that under sufficiently large amplitude disturbance, RR could transit
to unstable MR. Previous studies about dynamic transition (c.f. Kudyavtev et
al. 2002; Mouton \& Hornung 2007; Li, Gao \& Wu 2011) assumed implicitly $%
H_{R}>H_{R,\min }^{(MR)}$, so dynamic transition leads to stable MR. Here,
we indeed observe, through numerical simulation with density perturbation,
for the particular condition with $M_{0}=4$, $\theta _{w}=25^{o}$ and $%
H_{R}=0.328H_{A}$.

Numerical simulation shows various types of dynamic transition and displays
various complex shock interaction structure during dynamic transition within
the dual solution stability gap. One type is direct dynamic transition for
which the transition goes from RR to MR to unstart flow. The other type is
inverted dynamic transition, for which RR transits to MR but then transit
back to RR. Complex flow structures, such as hybrid MR --- type VI shock
interference, and double MR --- MR, are found to exist during the dynamic
transition, depending on how we provide the disturbance.

\textbf{Acknowledgement}.

Declaration of interests. The authors report no conflict of interest.

\end{document}